\documentclass{aa}

\usepackage{graphicx}
\usepackage[varg]{txfonts}
\usepackage{natbib}
\bibpunct{(}{)}{;}{a}{}{,}
\newcommand{\diff}{\mathop{}\!\mathrm{d}}

\begin{document}

  \title{How do giant planetary cores shape the dust disk?}

  \subtitle{HL Tau system}

  \author{Giovanni Picogna, Wilhelm Kley}

  \institute{Institut f\"{u}r Astronomie und Astrophysik, Kepler Center,
    Universit\"{a}t T\"{u}bingen, Auf der Morgenstelle 10, 72076 T\"{u}bingen,
    Germany\\
    \email{giovanni.picogna@uni-tuebingen.de}}

  \date{Received \today / Accepted \today}

  %\thanks{}

  \abstract
  %Context
  {We are observing, thanks to ALMA, the dust distribution in the
  region of active planet formation around young stars.
  This is a powerful tool to connect observations with theoretical models
  and improve our understandings of the processes at play.}
  %Aims
  {We want to test how a multi-planetary system shapes its
  birth disk and study the influence of the planetary masses and particle sizes
  on the final dust distribution.
  Moreover, we apply our model to the HL Tau system in order to obtain some
  insights on the physical parameters of the planets that are able to create
  the observed features.}
  %Methods
  {We follow the evolution of a population of dust particles, treated as
  Lagrangian particles, in two-dimensional, locally isothermal disks where two
  equal mass planets are present.
  The planets are kept in fixed orbits and they do not accrete mass.}
  %Results
  {The outer planet plays a major role removing the dust particles in the
  co-orbital region of the inner planet and forming a particle ring which
  promotes the development of vortices respect to the single planetary case.
  The ring and gaps width depends strongly on the planetary mass and particle
  stopping times, and for the more massive cases the ring clumps in few stable
  points that are able to collect a high mass fraction.
  The features observed in the HL Tau system can be explained through the
  presence of several massive cores that shape the dust disk, where
  the inner planet(s) should have a mass on the order of
  $0.07\,M_\mathrm{Jup}$ and the outer one(s) on the order of
  $0.35\,M_\mathrm{Jup}$.
  These values can be significantly lower if the disk mass turns out to be less
  than previously estimated.
  Decreasing the disk mass by a factor 10 we obtain similar gap widths for
  planets with a mass of $10\,M_\oplus$ and $20\,M_\oplus$ respectively.
  Although the particle gaps are prominent, the expected gaseous gaps
  would be barely visible.}
  %Conclusions
  {}


  \maketitle

  %%%%%%%%%%%%%%%%%%%%%%
  \section{Introduction}\label{sec:intro}
  %%%%%%%%%%%%%%%%%%%%%%
  The planetary cores of giant planets form on a timescale
  $\sim 1\,\mbox{Myr}$.
  In this relatively short time-span, a huge number of processes takes
  place, allowing a swarm of small dust particles to grow several order of
  magnitudes in size and mass, before the gas disk removal.
  Until now, the only observational constraints that we had in order to test
  planet formation models were the gas and dust emissions from
  protoplanetary nebulae on large scales (on the order of $\sim100\,\mbox{au}$)
  and the final stage of planet formation through the detection of full-fledged
  planetary systems.
  Thanks to the recent advent of a new generation of radio telescopes, like
  the Atacama Large Millimeter Array (ALMA), we are starting to get some
  pristine images of the formation process itself, resolving the dust component
  of protoplanetary disks in the region of active planet formation around young
  stars.

  An outstanding example of this giant leap in the observational data is the
  young HL Tau system, imaged by ALMA in Bands $3$, $6$, and $7$
  (respectively at wavelengths $2.9$, $1.3$ and $0.87\,\mbox{mm}$) with a
  spatial resolution up to $3.5\,\mbox{au}$.
  Several features can be seen in the young protoplanetary disk, but the most
  striking is the presence of several axysimmetric rings in the $\mbox{mm}$
  dust disk~\citep{Partnership2014}.
  Although different mechanisms can be responsible of the observed features
  \citep{Flock2015,Zhang2015}, the most straightforward explanation for the ring
  formation, in the sense that others mechanism require
  specific initial conditions that reduce their general applicability, is the
  presence of several planetary cores that grow in their birth
  disk and shape its dust content.
  Indeed, in order to have a particle concentration at a particular region, we
  need a steep pressure gradient in the gaseous disk which can trap particles,
  `sufficiently' decoupled from the gas, by changing its migration direction.
  A long-lived high-pressure region can be created even by a small mass
  planets \citep{Paardekooper2004}, which can effectively carve a deep dust gap
  and concentrate particles at the gap edges and at corotation in tadpole orbits
  \citep{Paardekooper2007, Fouchet2007}.

  The aim of this paper is to test how two giant planetary cores shape the
  dust disk in which they are born, implementing a particle population in the 2D
  hydro code \textsc{fargo}~\citep{Masset2000}, and study the influence of the
  planetary masses and particle sizes on the final disk distribution.
  Moreover, we apply our model to the HL Tau system in order to obtain some
  insights on the physical parameters of the planets creating the observed
  features.

  This paper is organised as follows.
  In Section~\ref{sec:model} we discuss under which physical conditions a planet
  is capable of opening a gap in the dust and gaseous disk in order to define
  the important physical scales for our model.
  In Section~\ref{sec:drag} we define the model adopted for the gas drag.
  Then, the setup of our simulations is explained in
  Section~\ref{sec:initsetup}, and the main results are outlined in
  Section~\ref{sec:res}.
  Finally, in Section~\ref{sec:disc} we discuss our results and their
  implications and limitation, while the major outcomes are highlighted in
  Section~\ref{sec:conc}.

  %%%%%%%%%%%%%%%%%%%%%%
  \section{Background}\label{sec:model}
  %%%%%%%%%%%%%%%%%%%%%%
  In order to set up our model, we need first to determine what is the
  minimum mass of a planetary core that is able to open up a gap in the gaseous
  and dust disk, for a given set of disk parameters, and the relative opening
  timescale.
  In particular, we want to understand the influence of the different physical
  processes modelled on the outcome of the simulation.

    %---------------------
    \subsection{Theory of gap formation}\label{sec:gap}
    %---------------------
    The theory of gap formation in gaseous disks has been studied
    extensively in the past, and there are a set of general criteria that a
    planet must fulfil in order to carve a gap.
    However, the possibility to open a gap in the dust disk is more complicated,
    since it depends strongly on the coupling between the dust and the gaseous
    media, and only recently this problem has been tackled.

      %---------------------
      \subsubsection{Gaseous gap}
      %---------------------
      The torque exchange between the disk and the planet adds angular momentum
      to the outer disk regions and removes it from the inner ones.
      As a result, the disk structure is modified in the regions close to the
      planet location and, given a minimum core mass and enough time, a gap
      develops.

      The time scale needed to open a gap of half width $x_\mathrm{s}$ can be
      crudely estimated from the impulse approximation \citep{Lin1979}.
      The total torque acting on a planet of mass $M_\mathrm{p}$ and semi-major
      axis $a$ due to its interaction with the outer disk of surface density
      $\Sigma$ is \citep{Lin1979,Papaloizou2006}
      \begin{equation}
        \label{eq:torqPlan}
        \frac{dJ}{dt}=-\frac{8}{27}\frac{G^2M_\mathrm{p}^2 a\Sigma}
        {\Omega_\mathrm{p}^2{x_\mathrm{s}}^3}
      \end{equation}
      where $\Omega_\mathrm{p}=\sqrt{GM_\star/a^3}$ is the planet orbital
      frequency around a star of mass $M_\star$.
      The angular momentum that must be added to the disk in order to remove the
      gas inside the gap is
      \begin{equation}
        \label{eq:AngMomRem}
        \Delta J=2\pi a x_\mathrm{s} \Sigma\frac{dl}{dr}\biggr|_{a}x_\mathrm{s}
      \end{equation}
      where $l=\sqrt{GM_\star r}$ is the gas specific angular momentum.
      Thus, the gap opening time can be estimated as
      \begin{align}
        \label{eq:topen}
        t_\mathrm{open} &= \frac{\Delta J}{|dJ/dt|} =
        \frac{27}{8}\pi\frac{1}{q^2\sqrt{GM_\star a}}
        \frac{{x_\mathrm{s}}^5}{a^3} \\
        &\simeq 33.8\,{\left(\frac{h}{0.05}\right)}^{5}
        {\left(\frac{q}{1.25*10^{-4}}\right)}^{-2} P_\mathrm{p}
      \end{align}
      where $q=M_\mathrm{p}/M_\star$ is the planet to star mass ratio, and we
      assumed that the minimum half width of the gap should be $x_\mathrm{s}=H$,
      where $H$ is the effective disk thickness, and $h=H/r$ is the normalised
      disk scale height.
      All values are evaluated at the planet location and the final estimate of
      the opening time is given in units of the planet orbital time
      $P_\mathrm{p}$.
      Although this is a crude estimate, it has been shown with more
      detailed descriptions that the order of magnitude obtained is correct.

      Based on this criterion, given enough time, even a small core can open a
      gap in an inviscid disk.
      But, we need to quantify the magnitude of the competing factors that act
      to prevent or promote its development, in order to obtain a better
      estimate of the gap opening time scale, and the minimum mass ratio.

        \paragraph{Thermal condition ---}
        The assumption made for the minimum gap half width in
        eq.~(\ref{eq:topen}) is necessary to allow non-linear dissipation of
        waves generated by the planet \citep{Korycansky1996} and to avoid
        dynamical instabilities at the planet location \citep{Papaloizou1984},
        which are necessary conditions to clear the regions close to the planet
        location.
        This condition, called thermal condition, translates into a first
        criterion for open up a gap
        \begin{equation}
          \label{eq:gapTherm}
          x_\mathrm{s}=1.16a\sqrt{\frac{q}{h}}\geq H = ha
        \end{equation}
        for a 2D disk \citep{Masset2006}, which correspond to a minimum planet
        to star mass ratio of
        \begin{equation}
          \label{eq:gapqTherm}
          q_\mathrm{th} \simeq h^3 = 1.25*10^{-4}\,
          {\left(\frac{h}{0.05}\right)}^3
        \end{equation}
        and a related thermal mass
        \begin{equation}
          \label{eq:gapMTherm}
          M_\mathrm{th} \simeq M_\star h^3 = 1.25*10^{-4} M_\star
        \end{equation}

        \paragraph{Viscous condition ---}
        The viscous diffusion acts to smooth out sharp radial gradients in the
        disk surface density, preventing the gap clearing mechanism.
        The time needed by viscous forces to close up a gap of width
        $x_\mathrm{s}$ is given by the diffusion timescale for a viscous fluid,
        which can be derived directly from the Navier-Stokes equation in
        cylindrical polar coordinates \citep[see e.g.][]{Armitage2010}
        \begin{equation}
          \label{eq:tvisc}
          t_\mathrm{visc}=\frac{x_\mathrm{s}^2}{\nu}\simeq 39.8\,
          {\left(\frac{\alpha}{0.004}\right)}^{-1} P_\mathrm{p}
        \end{equation}
        where $\nu$ is the kinematic viscosity, and
        $\alpha=\nu\Omega/c_\mathrm{s}^2$ is the Shakura-Sunyaev parameter that
        measures the efficiency of angular momentum transport due to turbulence.

        The minimum mass $q_\mathrm{visc}$ needed to open a gap in a viscous
        disk is obtain by comparing the opening time due to the torque
        interaction, eq.~(\ref{eq:topen}), with the closing time owing to
        viscous stress eq.~(\ref{eq:tvisc}) \citep{Lin1986,Lin1993}
        \begin{equation}
          \label{eq:gapVisc}
          q_\mathrm{visc} \simeq
          {\left(\frac{27}{8}\pi\right)}^{1/2}\alpha^{1/2}h^{5/2} \simeq
          1.15*10^{-4}{\left(\frac{\alpha}{0.004}\right)}^{1/2}
          {\left(\frac{h}{0.05}\right)}^{5/2}
        \end{equation}
        Thus, for the parameters chosen, the viscous condition is very similar
        to the thermal condition.

        \paragraph{Generalised condition ---}
        A more general semi-analytic criterion, which takes into account the
        balance between pressure, gravitational, and viscous torques for a
        planet on a fixed circular orbit has also been
        derived~\citep{Lin1993,Crida2006}
        \begin{equation}
          \label{eq:gapCrida}
          \frac{3}{4}\frac{H}{r_\mathrm{H}}+\frac{50}{q\ R_\mathrm{e}} < 1
        \end{equation}
        where $r_\mathrm{H}=a{(q/3)}^{1/3}$ is the Hill radius and
        $R_\mathrm{e}$ is the Reynolds number.
        From the previous relation, and plugging in the parameters used in our
        analysis we found that the minimum mass ratio is $q\simeq10^{-3}$.
        However, this criterion was derived for low planetary masses in low
        viscosity disks, and the behaviour might be considerably different
        changing those parameters.
        Moreover, this condition defines a gap as a drop of the mass density to
        $10\%$ of the unperturbed density at the planet’s location, but even a
        less dramatic depletion of mass affect planet-disk interaction and could
        be potentially detected.

      %---------------------
      \subsubsection{Dust gap}
      %---------------------
      In order to create a gap in the dust disk we need a radial pressure
      structure induced by the planet.
      Indeed, also a very shallow gap in the gas will change the drift speed of
      the dust particles significantly \citep{Whipple1972,Weidenschilling1977}
      favouring the formation of a particle gap.
      Thus, the minimum mass needed to open up a gap in the dust disk is a
      fraction of the one needed to clear a gap in the gas.
      ~\cite{Paardekooper2004,Paardekooper2006} performed extensive 2D
      simulations, treating the dust as a pressure-less fluid (which is a good
      approximation for tightly coupled particles), and they found that a
      planet more massive than $0.05 M_\mathrm{Jup}=0.38\,M_\mathrm{th}$ can
      open a gap in a mm size disk.
      Furthermore the dust gap opening time for this lower mass case was
      $50\,P_\mathrm{p}$, which is about half the timescale to open a gas gap.
      Previous studies found also a clear dependence of the gap width with the
      core mass, where larger planets open wider
      gaps~\citep{Paardekooper2006,Zhu2014}.
      Finally, there are some contrasting results regarding the dependence
      of the gap width respect to the particle
      size~\citep{Paardekooper2006}, but it seems that for less coupled
      particles the gap is wider for larger particles~\citep{Fouchet2007}.

	%---------------------
	\section{Model}\label{sec:drag}
	%---------------------
  Solid particles and gaseous molecules exchange momentum due to drag forces,
  that depend strongly on the condition of the gas and on shape, size and
  velocity of the particle.
  For sake of simplicity we limit ourselves to spherical particles.
  The drag force acts always in a direction opposite to the
  relative velocity.

  The regime that describes a particular system is defined by any two of
  three non-dimensional parameters.
  The Knudsen number $K = \lambda/s$ is the ratio of the two major
  length scales of the system: the mean free path of the gas molecules
  $\lambda$ and the particle size $s$.
  The Mach number $M = v_\mathrm{r}/c_\mathrm{s}$ is the ratio between the
  relative velocity between particles and gas $\mathbf{v}_\mathrm{r}$, and the
  gas sound speed $c_\mathrm{s}$.
  Finally, the Reynolds number $R_\mathrm{e}$ is related to different physical
  properties of the particle and gaseous media
  \begin{equation}
    R_\mathrm{e} = \frac{2 v_\mathrm{r}s}{\nu_\mathrm{m}}
  \end{equation}
  where $\nu_\mathrm{m}$ is the gas molecular viscosity, defined as
  \begin{equation}
    \nu_\mathrm{m} = \frac{1}{3}{\left(\frac{m_0\bar{v}_\mathrm{th}}{\sigma}
      \right)}
  \end{equation}
  where $m_0$ and $\bar{v}_\mathrm{th}$ are the mass and mean thermal velocity
  of the gas molecules, and $\sigma$ is their collisional cross section.

  There are two main regimes of the drag forces that we are going to study.

    %---------------------
    \subsection{Stokes regime}
    %---------------------
    For small Knudsen number, the particle experience the gas as a fluid.
    The drag force of a viscous medium with density $\rho_\mathrm{g}
    (\mathbf{r}_\mathrm{p})$ acting on a spherical dust particle with radius $s$
    can be modelled as \citep{Landau1959}
    \begin{equation}
      \mathbf{F}_\mathrm{D,S}=-\frac{1}{2}C_\mathrm{D}\pi s^2\rho_\mathrm{g}
      (\mathbf{r}_\mathrm{p}) v_\mathrm{r} \mathbf{v}_\mathrm{r}
    \end{equation}
    where the drag coefficient $C_\mathrm{D}$ is defined for the various regimes
    described above as \citep{Whipple1972,Weidenschilling1977}
    \begin{equation}
      C_\mathrm{D} \simeq
      \begin{cases}
        24\,{R_\mathrm{e}}^{-1} & R_\mathrm{e} < 1 \\
        24\,{R_\mathrm{e}}^{-0.6} & 1 < R_\mathrm{e} < 800 \\
        0.44 & R_\mathrm{e} > 800
      \end{cases}
    \end{equation}
    for low Mach numbers, however, for our choice of the parameter space
    high values are not expected in this regime.

    %---------------------
    \subsection{Epstein regime}
    %---------------------
    For low Knudsen numbers, the interaction between particles and single gas
    molecule becomes important.
    It can be modelled as \citep{Epstein1923}
    \begin{equation}
      \mathbf{F}_\mathrm{D,E}= -\frac{4}{3}\pi\rho_\mathrm{g}
      (\mathbf{r}_\mathrm{p})s^2\bar{v}_\mathrm{t}
      \mathbf{v}_\mathrm{r}
    \end{equation}

    %---------------------
    \subsection{General law}
    %---------------------
    The transition between the Epstein and Stokes regime occurs for a
    particle of size $s=9\lambda/4$ which in our case is a $m$ size particle in
    the inner disk, where the mean free path of the gas molecules is defined
    as \citep{Haghighipour2003}
    \begin{equation}
      \lambda = \frac{m_0}{\pi a_0^2\rho_\mathrm{g}(r)} =
      \frac{4.72*10^{-9}}{\rho_\mathrm{g}} [cm]
    \end{equation}
    for a molecular hydrogen particle with $a_0=1.5*10^8\,\mbox{cm}$.
    In order to model a broad range of particles sizes we adopt a linear
    combination of Stokes and Epstein regimes
    \citep{Supulver2000,Haghighipour2003}
    \begin{equation}
      \mathbf{F}_\mathrm{D} = (1-f)\mathbf{F}_\mathrm{D,E} +
      f\mathbf{F}_\mathrm{D,S}
    \end{equation}
    where the factor $f$ is related to the Knudsen number and is defined as:
    \begin{equation}
      f=\frac{s}{s+\lambda}=\frac{1}{1+\mathit{Kn}}
    \end{equation}

    %---------------------
    \subsection{Stopping time}
    %---------------------
    An important parameter to evaluate the strength of the drag force is the
    stopping time $t_\mathrm{s}$, which can be defined as
    \begin{equation}
      \mathbf{F}_\mathrm{D} = -\frac{m_\mathrm{s}}{t_\mathrm{s}}
      \mathbf{v}_\mathrm{r}
    \end{equation}
    where $m_\mathrm{s}$ is the mass of the single dust particle of density
    $\rho_\mathrm{s}$ and, in the Epstein regime, the stopping time can be
    expressed as
    \begin{equation}\label{eq:stop}
      t_\mathrm{s}=\frac{s\rho_\mathrm{s}}{\rho_\mathrm{g}\bar{v}_\mathrm{th}}
    \end{equation}
    It is useful also to derive an non-dimensional stopping time (or Stokes
    number) as
    \begin{equation}
      \tau_\mathrm{s}=\frac{s\rho_\mathrm{s}}{\rho_\mathrm{g}
      \bar{v}_\mathrm{th}}\Omega_\mathrm{K}(\mathbf{r})
    \end{equation}

  %---------------------
  \section{Setup}\label{sec:initsetup}
  %---------------------
  We used the \textsc{fargo} code \citep{Masset2000,Baruteau2008}, modified in
  order to take into account the evolution of partially decoupled particles.
  An infinitesimally thin disk around a star resembling the observed
  HL Tau system \citep{Kwon2011} is modelled.
  Thus, the vertically integrated versions of the hydrodynamical equations are
  solved in cylindrical coordinates ($r,\phi,z$), centred on the star where
  the disk lies in the equatorial plane ($z=0$).
  The resolution adopted in the main simulations is $256\times512$ with
  $250\,000$ dust particles for each size, although we tested also a case
  doubling the resolution in order to test whether our results were resolution
  dependent.

    %---------------------
    \subsection{Gas component}\label{par:gasdisk}
    %---------------------
    The initial disk is axysimmetric and it extends from $0.1$ to $4$ in code
    units, where the unit of length is $25\,\mbox{au}$.
    The gas moves with azimuthal velocity given by the Keplerian speed around a
    central star of mass $1$ ($0.55\,M_\odot$), corrected by the rotating
    velocity of the coordinate system.
    We assume no initial radial motion of the gas, since a thin Keplerian disk
    is radially in equilibrium as gravitational and centrifugal forces
    approximately balance because pressure effects are small.
    The initial surface density profile is given by
    \begin{equation}
      \label{eq:surfprof}
      \Sigma(r)=\Sigma_0\ r^{-1}
    \end{equation}
    where $\Sigma_0$ is the surface density at $r=1$ such as the total disk mass
    is equal to $0.24$ ($0.13\ M_\odot$) in order to match the value found by
    \citet{Kwon2011}.
    The disk is modelled with a locally isothermal equation of state, which
    keeps constant the initial temperature stratification throughout the whole
    simulation.
    We assumed a constant aspect ratio $H/r = 0.05$, that corresponds to a
    temperature profile
    \begin{equation}
      \label{eq:tempprof}
      T(r) = T_0\ r^{-1}
    \end{equation}
    We introduce a density floor of $\Sigma_\mathrm{floor} = 10^{-9}\ \Sigma_0$
    in order to avoid numerical issues.
    For the inner boundary we applied a zero-gradient outflow condition, while
    for the outer boundary we adopted a non-reflecting boundary condition.
    In addition, to maintain the initial disk structure in the outer parts of
    the disk we implemented a wave killing zone close to the boundary
    \citep{deVal-Borro2006},
    \begin{equation}\label{eq:dumping}
      \frac{d\xi}{dt}=-\frac{\xi-\xi_0}{\tau} {R(r)}^2
    \end{equation}
    where $\xi$ represent the radial velocity, angular velocity, and surface
    density.
    Those physical quantities are damped towards their initial values on a
    timescale given by $\tau$ and $R(r)$ is a linear ramp-function decreasing
    from $1$ to $0$ from $r=3.6$ to the outer radius of the computational
    domain.
    The details of the implementation of the boundary conditions can be found in
    \citet{Muller2012}.
    For the viscosity we adopt a constant $\alpha$ viscosity $\alpha = 0.004$.
    Furthermore, we discuss the gravitational stability of such initial
    configuration in Appendix~\ref{sec::stability}.

    %---------------------
    \subsection{Dust component}\label{par:dustdisk}
\begin{figure}
      \centering
      \includegraphics[width=.45\textwidth]{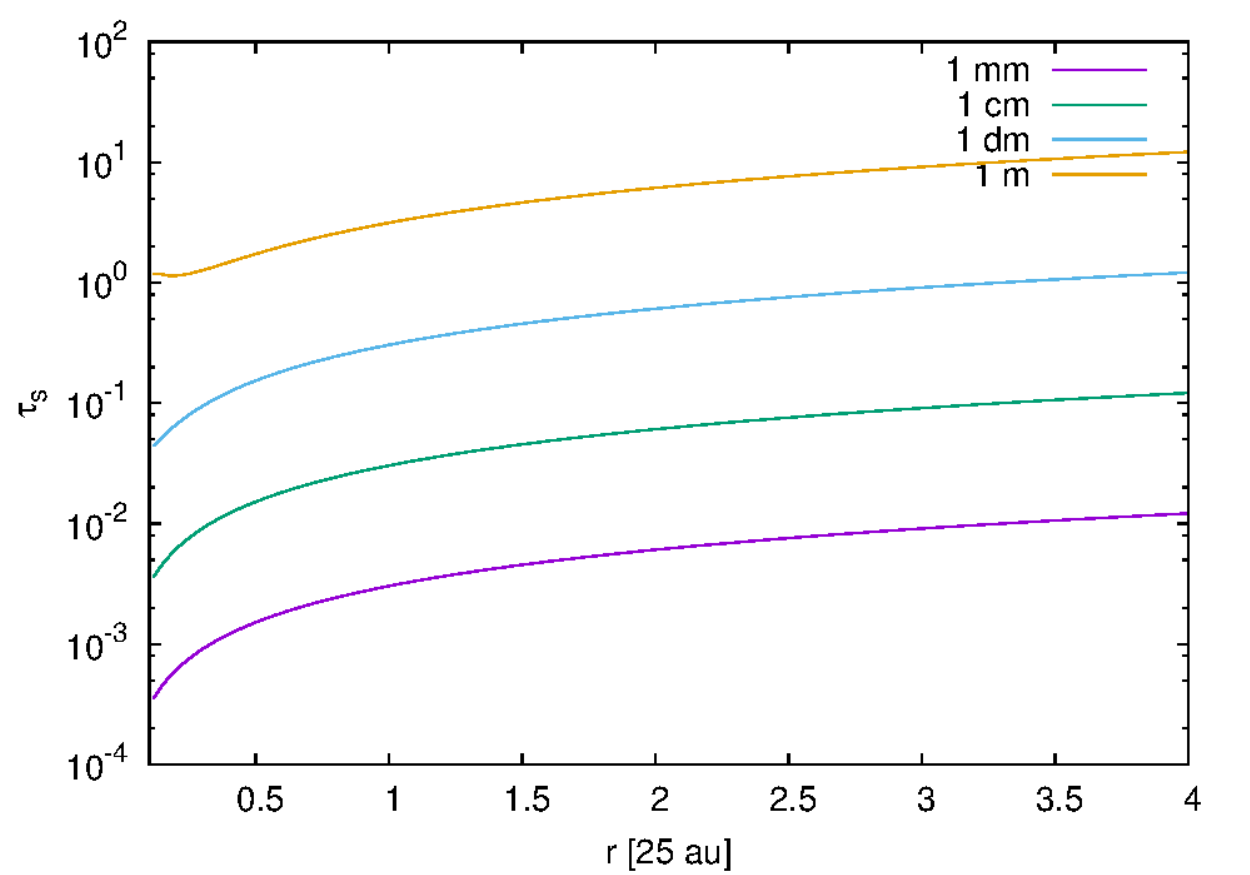}
      \caption{Stopping time at the beginning of the simulation for the
        different particle sizes modelled.\label{Fig:Ts}}
    \end{figure}

    %---------------------
    The solid fraction of the disk is modelled with $250\,000$ Lagrangian
    particles for each size considered.
    We study particles with sizes of $\mbox{mm},\mbox{cm},\mbox{dm},\mbox{m}$,
    and internal density $\rho_\mathrm{d}=2.6\ \mbox{g/cm}^3$.
    The initial surface density profile for the dust particles is flat
    \begin{equation}
      \Sigma_\mathrm{s}(r) = \Sigma_\mathrm{s,0}
    \end{equation}
    This choice was made in order to have a larger reservoir of particles in
    the outer disk, since at the beginning of the simulation the planets are
    slowly growing, thus they are unable to filtrate particles efficiently.
    The particles were introduced at the beginning of the simulation, and they
    are evolved with two different integrators, depending on their stopping
    times.
    Following the approach by~\cite{Zhu2014}, we adopted a semi-implicit
    Leapfrog-like (Drift-Kick-Drift) integrator in polar coordinates for larger
    particles, and a fully implicit integrator for particles well coupled to the
    gas.
    For the interested reader, we have added in Appendix~\ref{sec::integrators}
    the detailed implementation of the two integrators.
    In this work we do not take into account the back-reaction of the particle
    on the gaseous disk since we are interested only in the general structure of
    the dust disk and not on the evolution of dust clumps.
    Furthermore, for sake of simplicity and to speed up our simulations, we do
    not consider the effect of the disk self-gravity on the particle evolution.
    This could be in principle an important factor for the young and massive
    HL Tau system, although no asymmetric structures related to a
    gravitationally unstable disk are observed.
    Finally, we do not consider particle diffusion by disk turbulence, which
    could be important to prevent strong clumping of particles
    (Baruteau, private communication), and it will be the subject of a future
    study.
    In Figure~\ref{Fig:Ts} we plotted the stopping times calculated at the
    beginning of the simulations for the various particle species modelled.
    The smaller particles (cm, and mm-size) are strongly coupled to the gas in
    the whole domain, while dm-size particles approach a stopping time of order
    unity in the outer disk, and m particles in the inner part were we can see
    also a change in the profile due to the passage from the Epstein to Stokes
    drag regime.

\begin{figure}
      \centering
      \includegraphics[width=.45\textwidth]{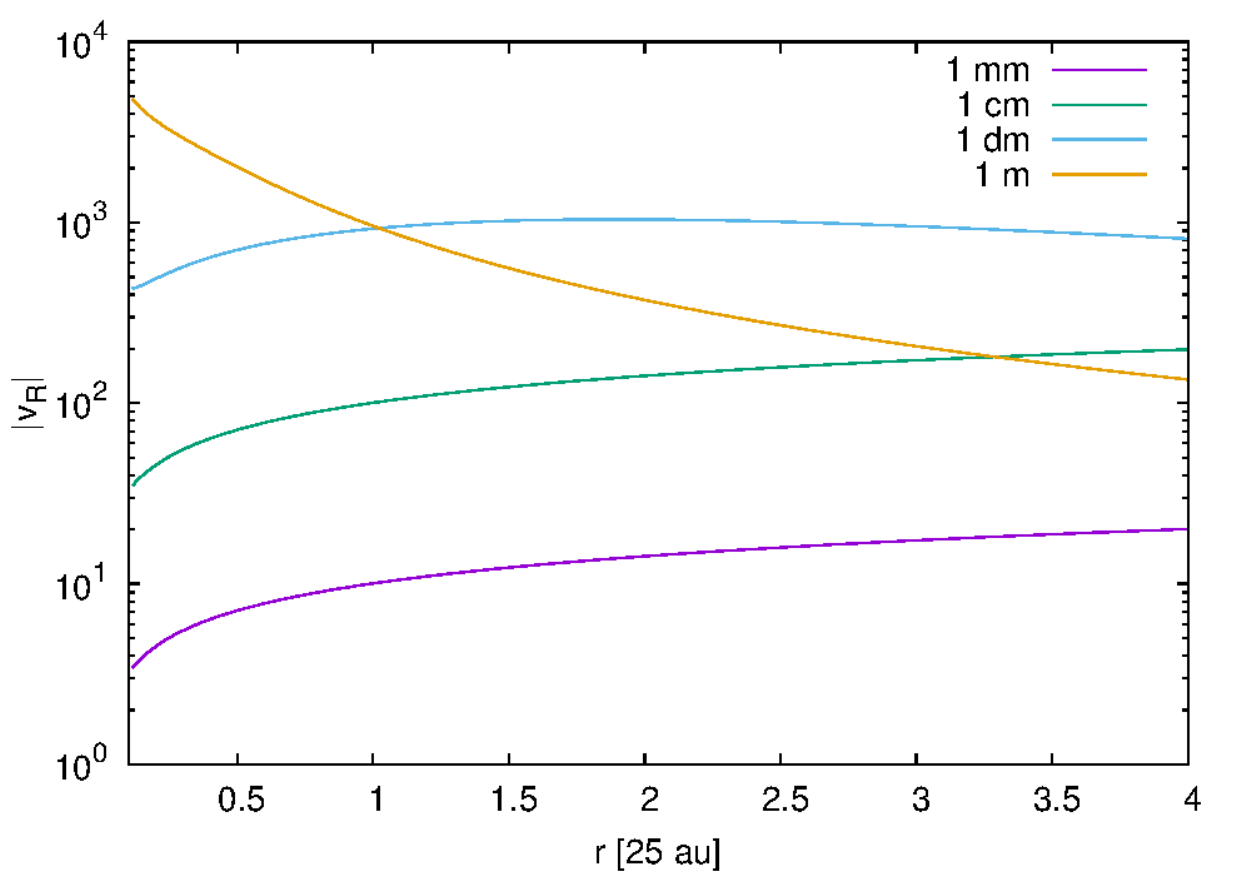}
      \caption{Radial drift velocity profile at the equilibrium for the
        different particle sizes modelled.\label{Fig:Vrad}}
    \end{figure}

    The transition between the Epstein and Stokes regime is clearly visible in
    Figure~\ref{Fig:Vrad} where the radial drift velocity at the equilibrium is
    plotted for the different particles sizes in the whole domain.
    As the particles approaches a stopping time of order unity their radial
    velocity grows, so the highest value is associated to the dm particles in
    the outer disk and the m-size particles in the inner parts.
    Furthermore, due to the transition between the two drag regimes, the profile
    of the curves rapidly changes from cm to m-size particles.
    We point out that when the planet start to clear a gap, the gas surface
    density inside it drops, and thus the stopping time of particles in
    horseshoe orbit can increase up to 2 orders of
    magnitude~\citep{Paardekooper2006}.
    The transition between the different drag regimes is then expected not only
    in the inner part of the disk but also near the planet co-orbital regions.

    %---------------------
    \subsection{Planets}\label{par:planets}
    %---------------------
    We embed two equal mass planets that orbit their parent star in circular
    orbits with semi-major axes $a_1=1$ and $a_2=2$.
    Their mass ranges from $1\,M_\mathrm{th}=0.07\,M_\mathrm{Jup}$ to
    $10\,M_\mathrm{th}=0.7\,M_\mathrm{Jup}$.
    Mass accretion is not allowed, and the planets do not feel the disk, so
    their orbital parameters remain fixed during the whole simulation.
    The gravitational potential of the planets is modelled through a Plummer
    type prescription, which takes into account the vertical extent of the disk
    and avoids the numerical issues related to a point-mass potential.
    We used a smoothing value of $\epsilon = 0.6\ H$ as this describes the
    vertically averaged forces very well \citep{Muller2012}.
    To prevent strong shock waves in the initial phase of the simulations the
    planetary core mass is increased slowly over $20$ orbits.
    Tab.~\ref{tab:sum} summarises the parameters of the standard model.
    \begin{table}
      \caption{Models}\label{tab:sum}
      \centering
      \begin{tabular}{c c}
        \hline\hline
        Parameter & Range \\
        \hline
        Planet mass [$M_\mathrm{th}$] & $1$,   $5$, $10$ \\
        Dust size [$\mbox{cm}$]       & $0.1$, $1$, $10$, $100$ \\
        \hline
      \end{tabular}
    \end{table}
    With these values the gaps are not expected to overlap since, even for the
    highest mass planet, from eq.~(\ref{eq:gapTherm}) we have
    $x_\mathrm{s}\simeq0.18$, thus
    \begin{equation}
      a_1+x_\mathrm{s} < a_2-x_\mathrm{s}
    \end{equation}
    The simulations were run for 600 orbital times of the inner planet, which
    corresponds to $\sim 200$ orbits of the outer planet, and to
    $\sim 10^5\,\mbox{yr}$, which is a consistent amount of time for a planetary
    system around a young star like HL Tauri which has only $10^6\,\mbox{yr}$.

  %%%%%%%%%%%%%%%%%%%%%%
  \section{Results}\label{sec:res}
  %%%%%%%%%%%%%%%%%%%%%%

    %---------------------
    \subsection{Massive core ($10 M_\mathrm{th}$)}\label{sec:res10}
    %---------------------
    Based on the criteria of gap opening reviewed in Section~\ref{sec:model},
    two $10 M_\mathrm{th}$ planets should open up rapidly a gap in the gas and
    dust disk.
    We study in detail the disk evolution for these massive cores, focusing on
    particle concentration which happen mainly at gap edges.

    In the following analysis the region of high surface density between the two
    planets is referred as the ring.
    It will have an inner and outer edge, which correspond to the outer edge of
    the inner gap and the inner edge of the outer gap, respectively.
    Furthermore, there will be the outer edge of the outer gap and the
    inner edge of the inner gap.
    The study of the various gap edges, where there is an abrupt change in the
    surface density profile, is important since those are potentially unstable
    regions where gas and particles could collect, changing the final surface
    density distribution of the disk.

      %---------------------
      \subsubsection{Gas distribution}
\begin{figure}
        \centering
        \includegraphics[width=.45\textwidth]{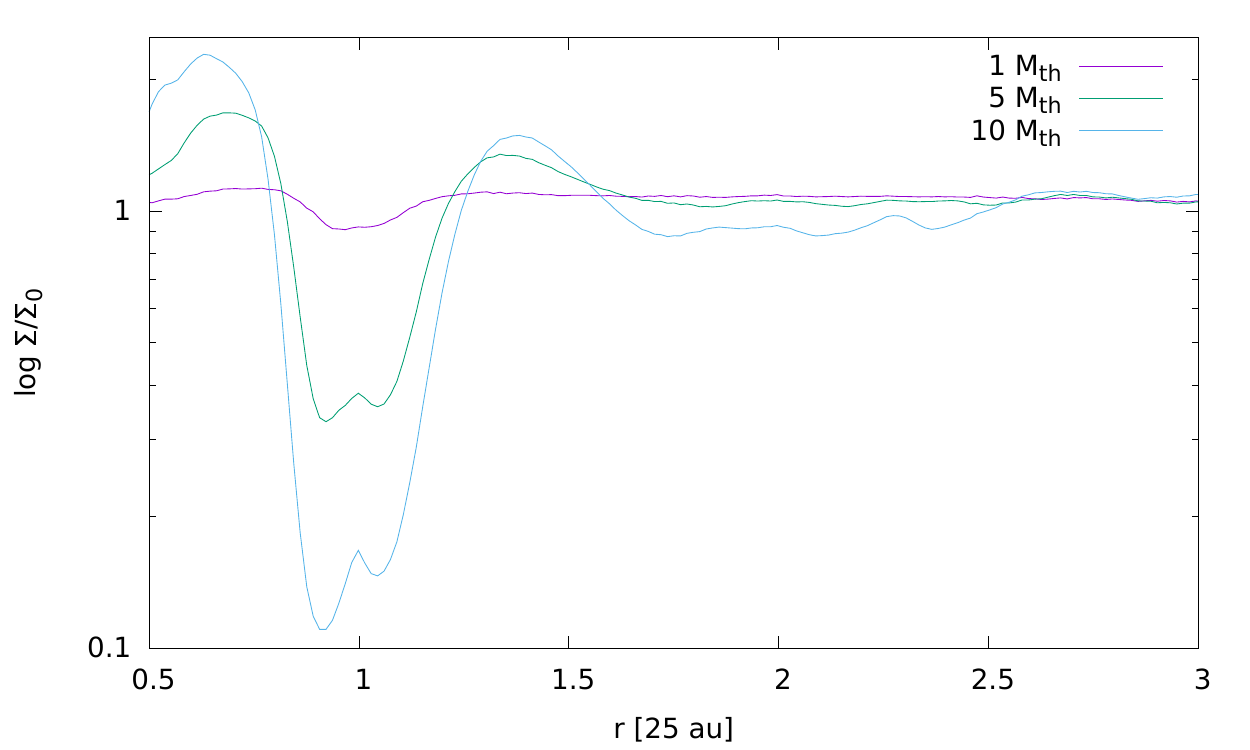}
        \includegraphics[width=.45\textwidth]{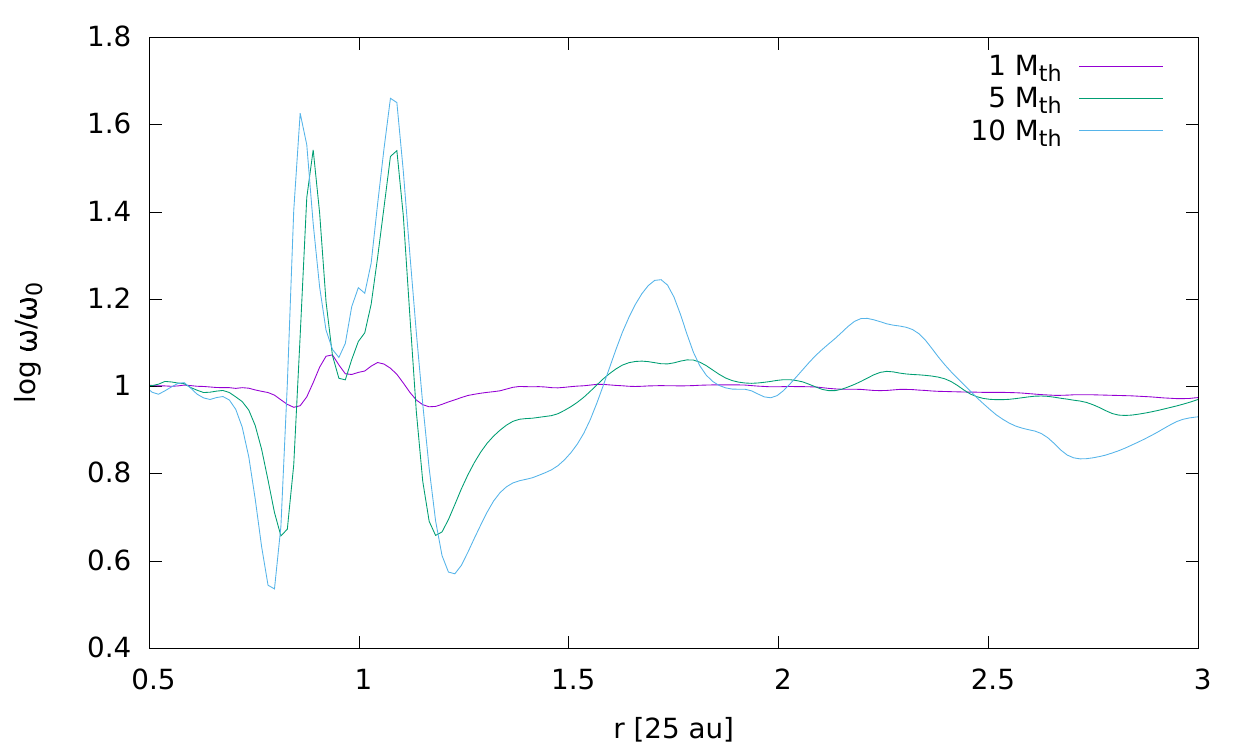}
        \caption{Gas surface density (top panel) and vorticity (bottom panel)
          profile after $100$ orbits.\label{Fig:Surf10}}
      \end{figure}

      %---------------------
      The inner planet has already opened a clear gap after 100 orbits
      (Figure~\ref{Fig:Surf10}~-~top panel) where, as expected by the general
      criterion of~\cite{Crida2006}, the surface density is an
      order of magnitude lower than its unperturbed value.
      Meanwhile, the outer planet is still opening its gap since it has a longer
      dynamical timescale.

\begin{figure}
        \centering
        \includegraphics[width=.45\textwidth]{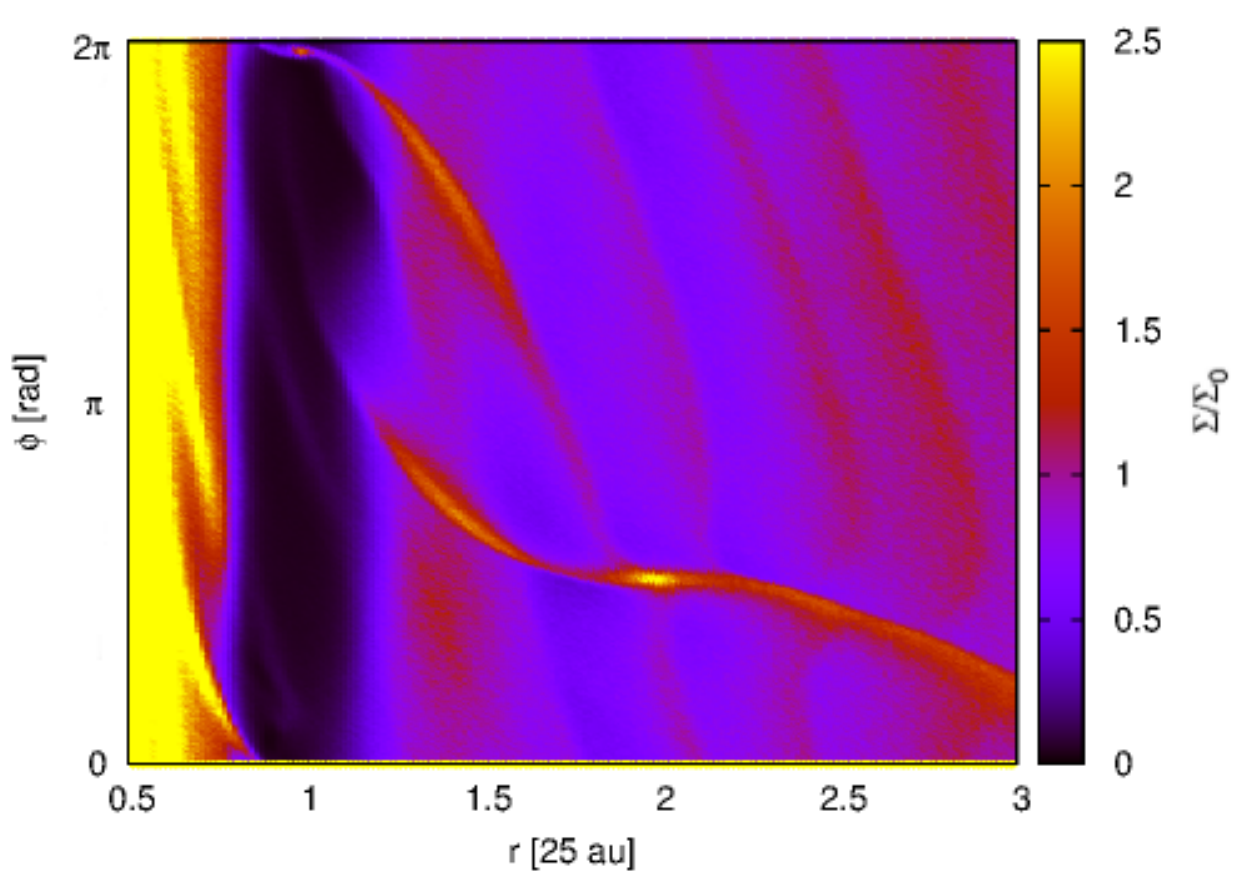}
        \includegraphics[width=.45\textwidth]{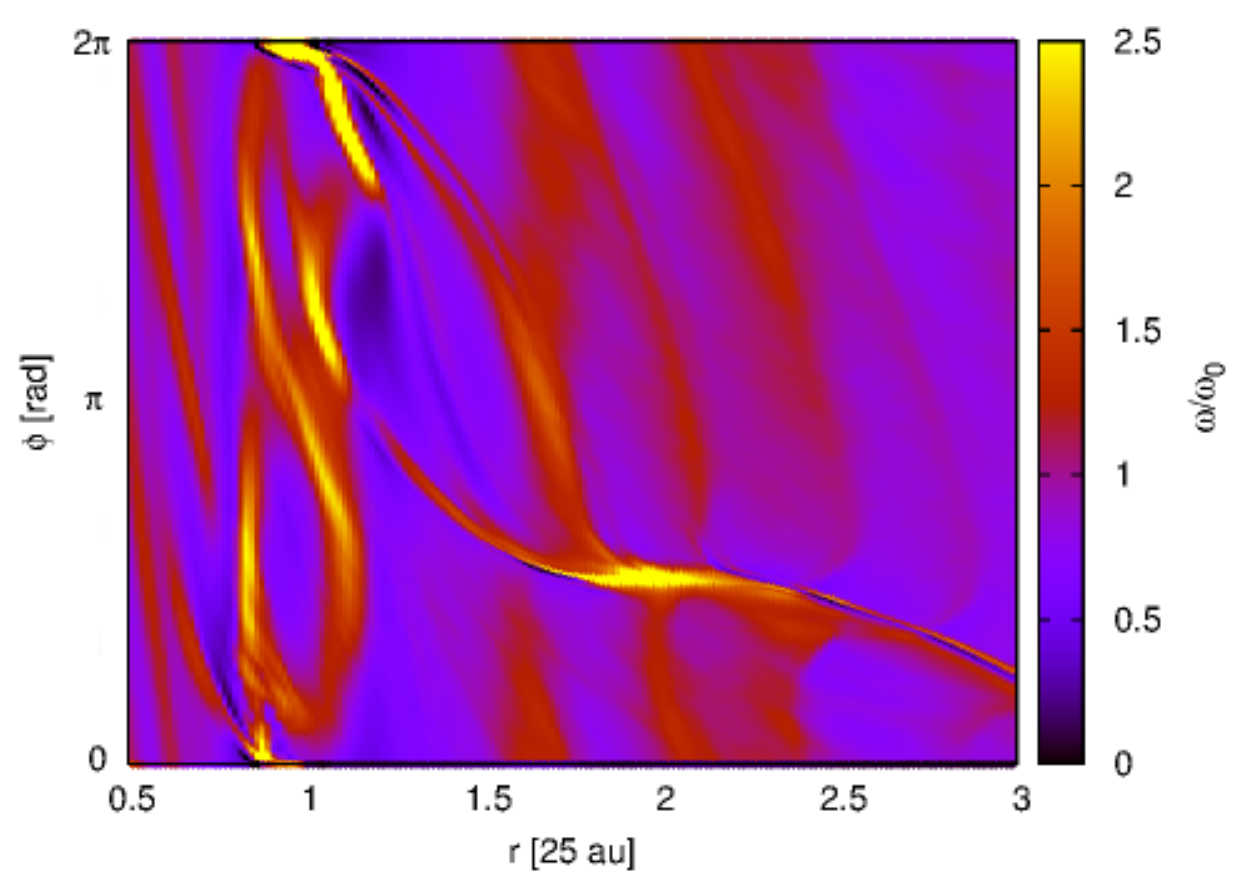}
        \caption{Gas surface density (top panel) and vorticity distribution
          (bottom panel) at the end of the simulation for the massive cores
          case.~\label{Fig:Surf2D10}}
      \end{figure}

      The steep profile in the surface density close to the ring inner edge can
      trigger a Rossby wave instability (RWI,~\cite{Li2001}).
      This instability gives rise to a growing non-axisymmetric perturbation
      consisting of anticyclonic vortices.
      A vortex is able to collect a high mass fraction and it can change
      significantly the final distribution of gas and dust in the disk.
      An important parameter to study, when considering the evolution of
      vortices is the gas vorticity, which is defined as
      \begin{equation}
        \omega_\mathrm{z}={(\nabla\times \mathbf{v})}_\mathrm{z}
      \end{equation}
      We show its profile in Figure~\ref{Fig:Surf10} (bottom panel)
      and its 2D distribution in Figure~\ref{Fig:Surf2D10} (bottom panel).
      Comparing the 2D distributions of vorticity and surface density
      (Figure~\ref{Fig:Surf2D10}), we see that vorticity peaks where
      the gap is deeper, and low vorticity regions appear at the centre
      of spiral arms created by the planets and close to the ring inner
      edge.

      The development of vortices due to the presence of a planet has been
      studied extensively.
      However, their evolution in a multi-planet system has not yet been
      addressed.
      From Figure~\ref{Fig:Surf2D10} we see that the outer planet perturbs
      substantially the co-orbital region of the inner one.
      There are two competing factors that need to be taken into account in
      order to estimate the lifetime of a vortex.
      On the one hand vortex formation is promoted by the enhanced surface
      density gradient at the ring location due to the combined action of both
      planets that push away the disk from their location.
      On the other hand, the periodic close encounters of the outer planet with
      the vortices enhance the eccentricity of the dust particles trap into it,
      favouring their escape and thus depleting the solid concentration inside
      the vortex.

      The capacity of collect particles by a vortex is closely linked to its
      orbital speed.
      If a vortex has a Keplerian orbital speed, then dust particles with
      the same orbital frequency will remain in the vortex for many orbits and
      slowly drift to its centre due to drag forces.
      On the other hand, a particle in a vortex orbiting with a non-Keplerian
      frequency will experience a Coriolis force in the Keplerian reference
      frame and, if the drag force is unable to counteract it, the particle will
      leave the vortex location~\citep{Youdin2010}.

\begin{figure}
        \centering
        \includegraphics[width=.45\textwidth]{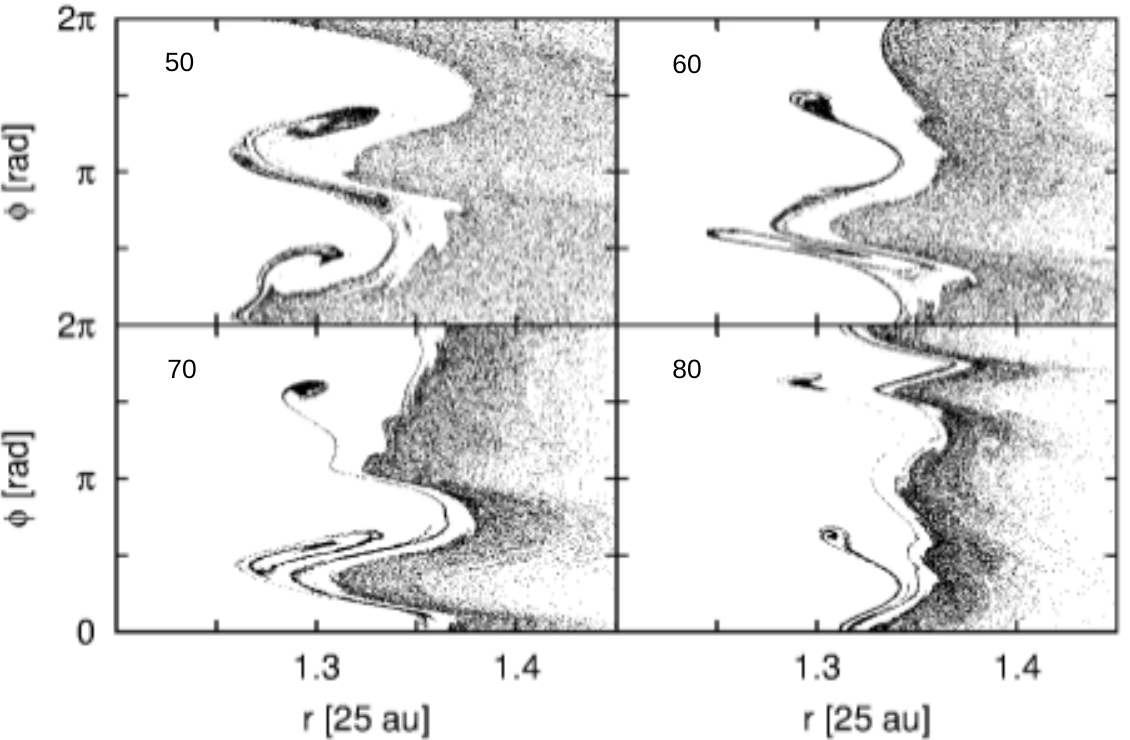}
        \caption{cm-sized particle distribution at different time-steps (50, 60,
          70, 80 orbital times) at the ring inner edge for the massive cores
          case.
          The evolution of two vortices in a frame co-moving with the disk at
          $1.3\,\mbox{au}$ is shown.
          The vortex centre orbits the central star with a velocity close to the
          background Keplerian speed, which promotes particle
          trapping inside the vortex.~\label{Fig:Vortevol}}
      \end{figure}

      The evolution of vortices can be also studied from the dynamics of coupled
      particles, which follow closely the gas dynamics.
      From the analysis of cm particles near the ring inner edge
      (Figure~\ref{Fig:Vortevol}), we can see that after 50 orbits two vortices
      are already visible and they last for several tens of orbits.
      The outer planet stretches periodically the vortices, and as a result they
      slowly shrink in size.

      In order to see if the vortices that develop in our simulation are capable
      of collecting a large fraction of particles, we plot in
      Figure~\ref{Fig:Vortevol} the vortices in a co-moving frame with the disk
      at $r=1.3$.
      The two vortices follow within a few percent the Keplerian speed, thus
      they are potentially able to trap a consistent fraction of particles.

\begin{figure}
        \centering
        \includegraphics[width=.45\textwidth]{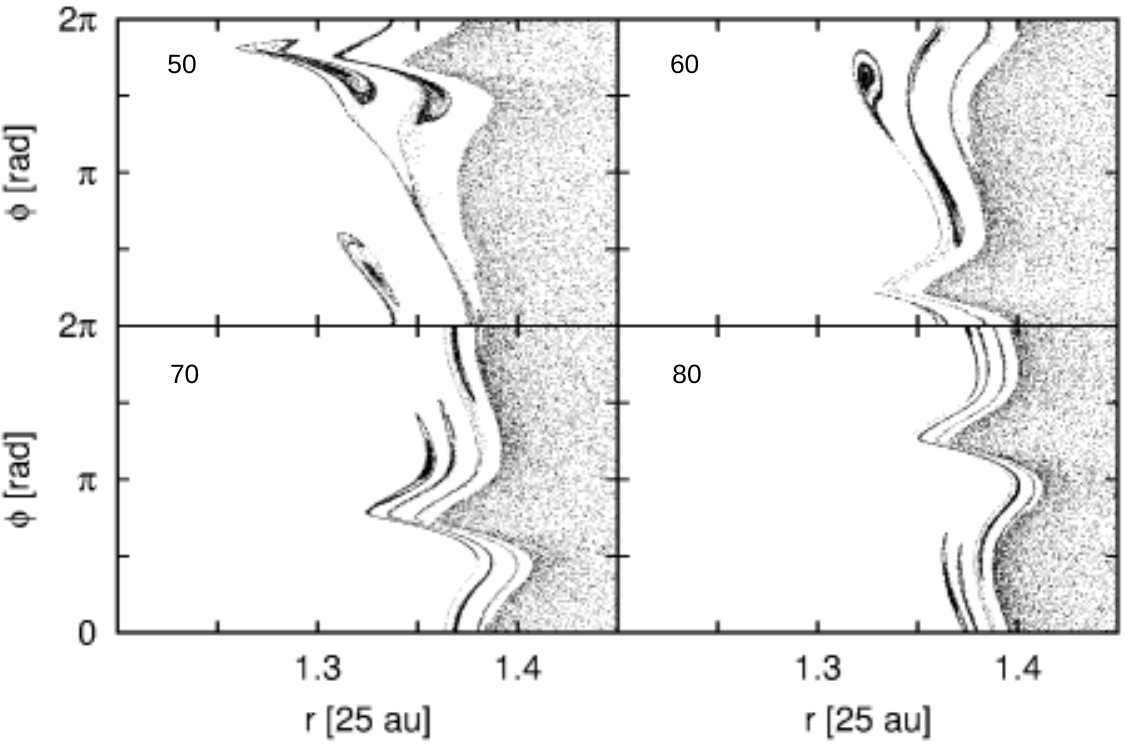}
        \caption{Same as Fig.~\ref{Fig:Vortevol} but for a single massive core
          at $r=1$.~\label{Fig:Comp1}}
      \end{figure}

      The influence of the outer planet on the development of RWI at the gap
      edge is studied by running a different model with only one massive planet
      ($M=10 M_\mathrm{th}$) at $r = 1.0$ (Figure~\ref{Fig:Comp1}).
      The particle concentration near the inner ring is weaker in the
      single core simulation.
      As a result, the vortices that form are less prominent and, although the
      perturbation to the ring inner edge is reduced respect to the dual core
      simulation, their lifetime is shorter.

      %---------------------
      \subsubsection{Particle distribution}
\begin{figure}
        \centering
        \includegraphics[width=.38\textwidth]{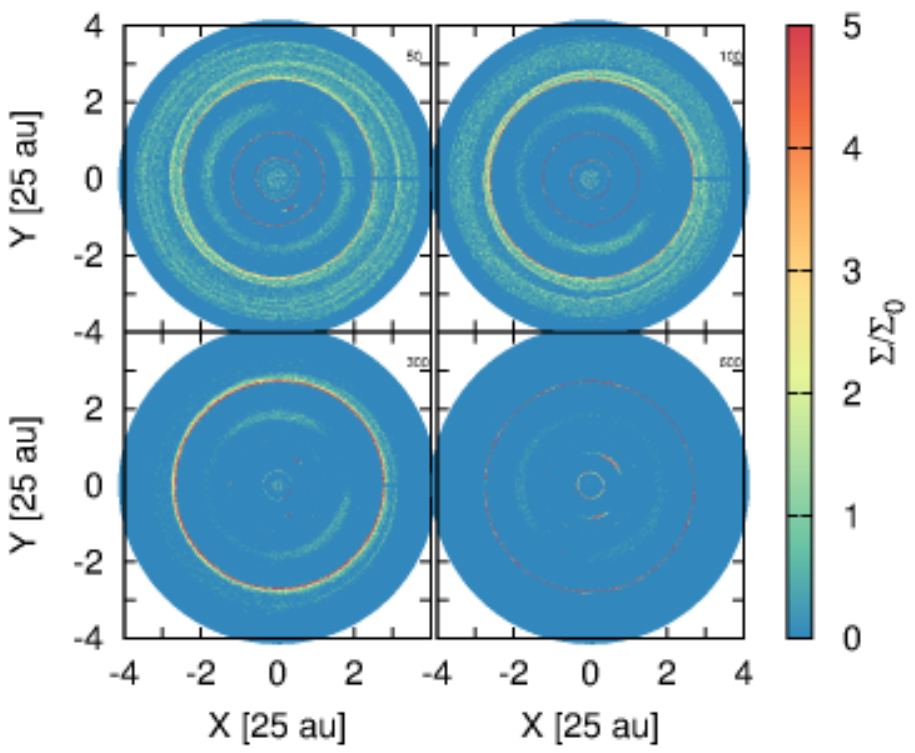}
        \includegraphics[width=.38\textwidth]{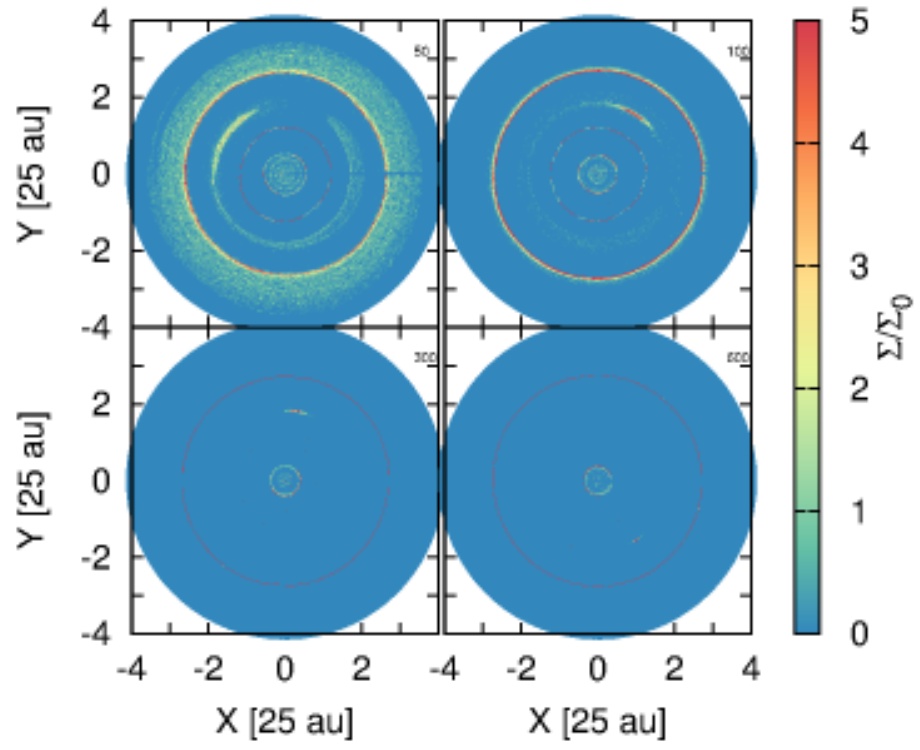}
        \includegraphics[width=.38\textwidth]{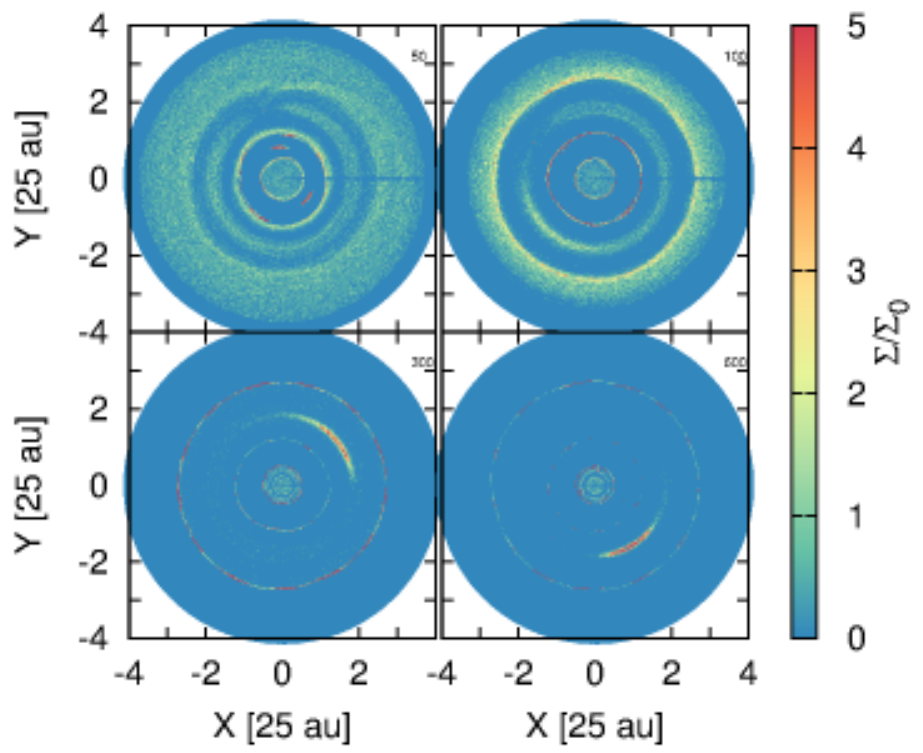}
        \includegraphics[width=.38\textwidth]{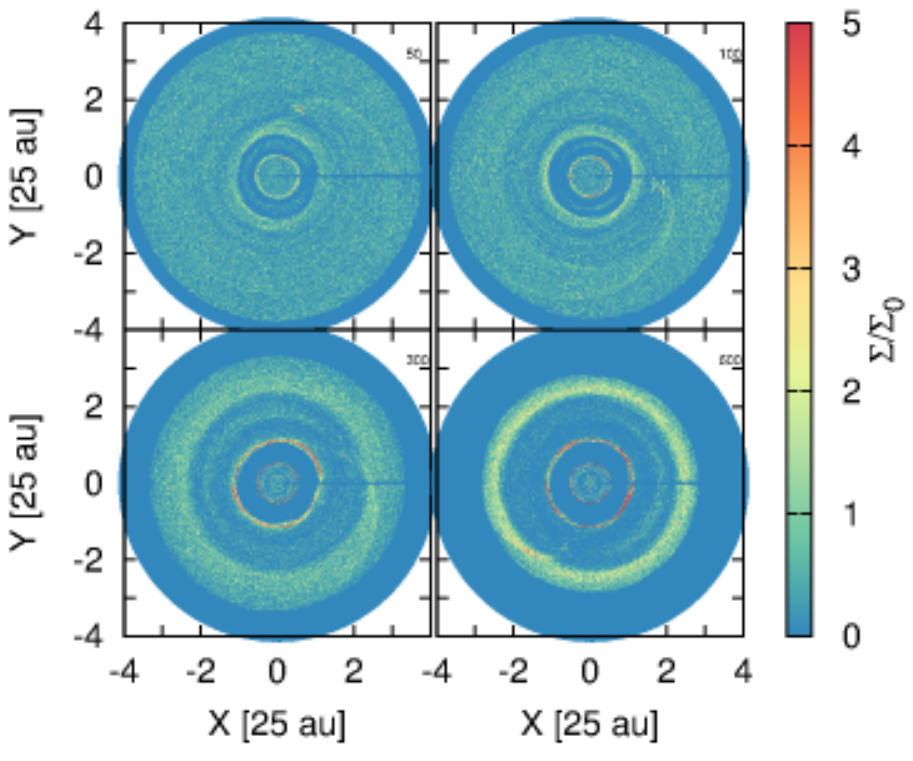}
        \caption{Dust normalised surface density distribution for the m (first
          panel), dm (second panel), cm (third panel), and mm-sized (fourth
          panel) particles disk and 2 equal mass $10\,M_\mathrm{th}$ cores at
          $r=1,2$ at 4 different times (50, 100, 300, 600 orbital times) for
          each case.~\label{Fig:Dust10}}
      \end{figure}

      %---------------------
      The evolution of the normalised surface density of the various dust
      species is shown in Figure~\ref{Fig:Dust10}.
      The inner and outer planets carve rapidly (first $50$ orbits) a particle
      gap, except for the most coupled particles simulated (mm-sized), which are
      cleared on a longer timescale for the outer planet
      (see Figure~\ref{Fig:Dust10}~-~fourth panel).
      As stated before, this behaviour is only due to the longer dynamical
      timescale of the outer planet, and it follows closely the evolution of the
      gas in the outer gap.

\begin{figure}
        \centering
        \includegraphics[width=.45\textwidth]{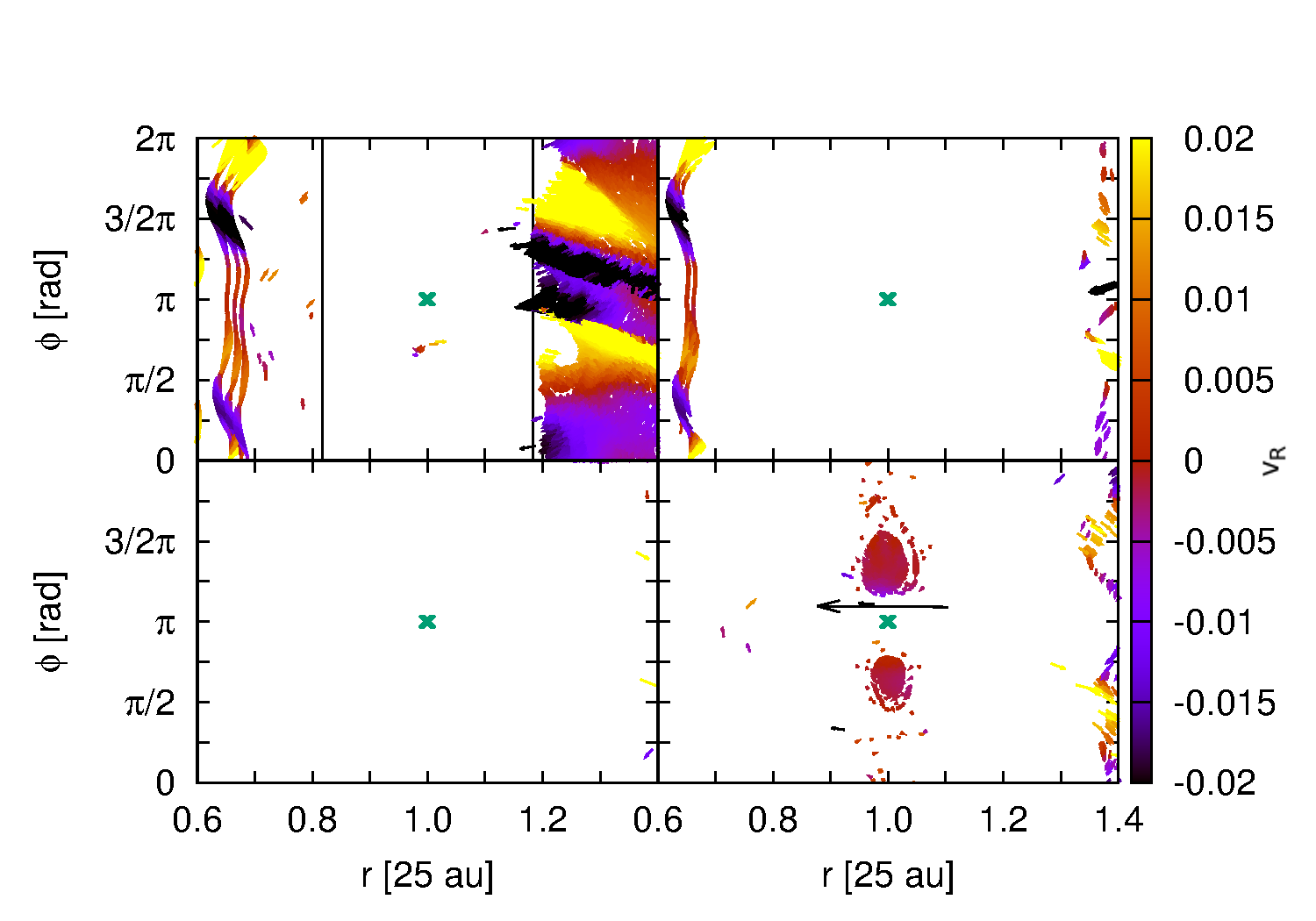}
        \caption{Particle distribution near the inner massive planetary core
          location for mm (top left), cm (top right), dm (bottom left) and m
          (bottom right) size particles at the end of the simulation.
          The velocity vectors of the particles respect to the planet are shown
          and the colour scale highlight the radial velocity
          component.~\label{Fig:Gap10m}}
      \end{figure}

      A significant fraction of particles clumps in the co-orbital region with
      the inner planet for several hundred periods.
      However, they are finally disrupted by the tidal interaction with the
      outer planet which excite their eccentricities, causing a close fly-by
      with the inner planet.
      The only particles which remain for the all simulated period in the
      co-orbital region are the m-size ones, although even they are perturbed
      and a significant mass exchange between the two Lagrangian points (L4 and
      L5) takes place (see Figure~\ref{Fig:Dust10}~-~first panel and
      Fig.~\ref{Fig:Gap10m}~-~bottom right panel).
      The outer planet is also able to keep a fraction of particles in the
      co-orbital region for longer timescale, though they are much more
      dispersed respect to the Lagrangian points.
      Moreover, in all simulations the L4 Lagrangian point is the most
      populated.

\begin{figure}
        \centering
        \includegraphics[width=.45\textwidth]{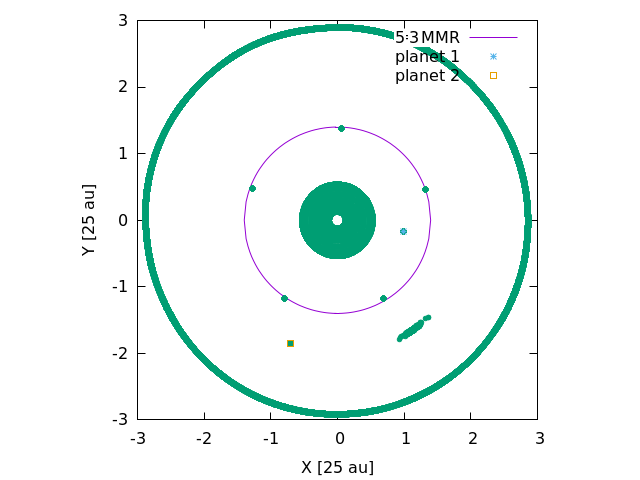}
        \caption{dm sized particle distribution at the end of the simulation of
          the two massive cores.
          The particle ring between the two planets shrinks with time until it
          clumps in few stable points that forms a pentagon-like
          structure with an orbit close to the 5:3 MMR with the inner
          planet at $r=1.4$.~\label{Fig:Clumping}}
      \end{figure}

      The ring, which forms between the two gaps, gets shortly very narrow and
      it stabilises in a position close to the 5:3 mean motion resonance (MMR)
      with the inner planet which seems a stable orbit.
      As shown in Figure~\ref{Fig:Clumping} the particles in the ring clump
      around $5$ symmetric points which gain a high mass.

      There are visible vortices in the particle distribution in the first
      hundred orbits for the cm size particles at the inner ring edge (third
      panel).
      Although these structures are prominent in the particle distribution of
      cm size dust, they are not visible in the other dust size distributions.
      The main reason is that for larger particles the ring get shortly very
      narrow and so there is no time for them to be trapped into the vortex.

      The velocity components of the particles shown in
      Figure~\ref{Fig:Gap10m} highlight the strong perturbations due
      to the spiral arm generated by the outer planet which affects mainly the
      most coupled particles and the gas.
      The bodies passing close to the planet location, as in the meter case
      shown in Figure~\ref{Fig:Gap10m} (bottom right panel), gain a high
      velocity component that is represented by the long black arrow.

      Finally, the particle distributions at the end of the simulation
      (see Figure~\ref{Fig:Gap10m}) show the dependence of gap width on
      particle size.
      The mm size particles (Fig.~\ref{Fig:Gap10m}~-~top left panel) reach a
      region closer to the planet location, where we overplot the minimum
      half-gap width as calculated from the eq.~(\ref{eq:gapTherm}), showing
      that the dynamics of the smallest particles follow closely that of the
      gas.
      The others particles have increasingly larger gaps.
      The dm particles (Fig.~\ref{Fig:Gap10m}~-~bottom left panel) have cleared
      almost completely the gap region since they have a stopping time closer
      to one near the planet location and thus their evolution is faster.

    %---------------------
    \subsection{Intermediate mass core ($5 M_\mathrm{th}$)}
    %---------------------
    The intermediate mass planet should still open up a gap in the gaseous disk,
    but on a much longer timescale, since from eq.~(\ref{eq:topen}) the
    opening time scales as $q^{-2}$.

      %---------------------
      \subsubsection{Gas distribution}
\begin{figure}
        \centering
        \includegraphics[width=.45\textwidth]{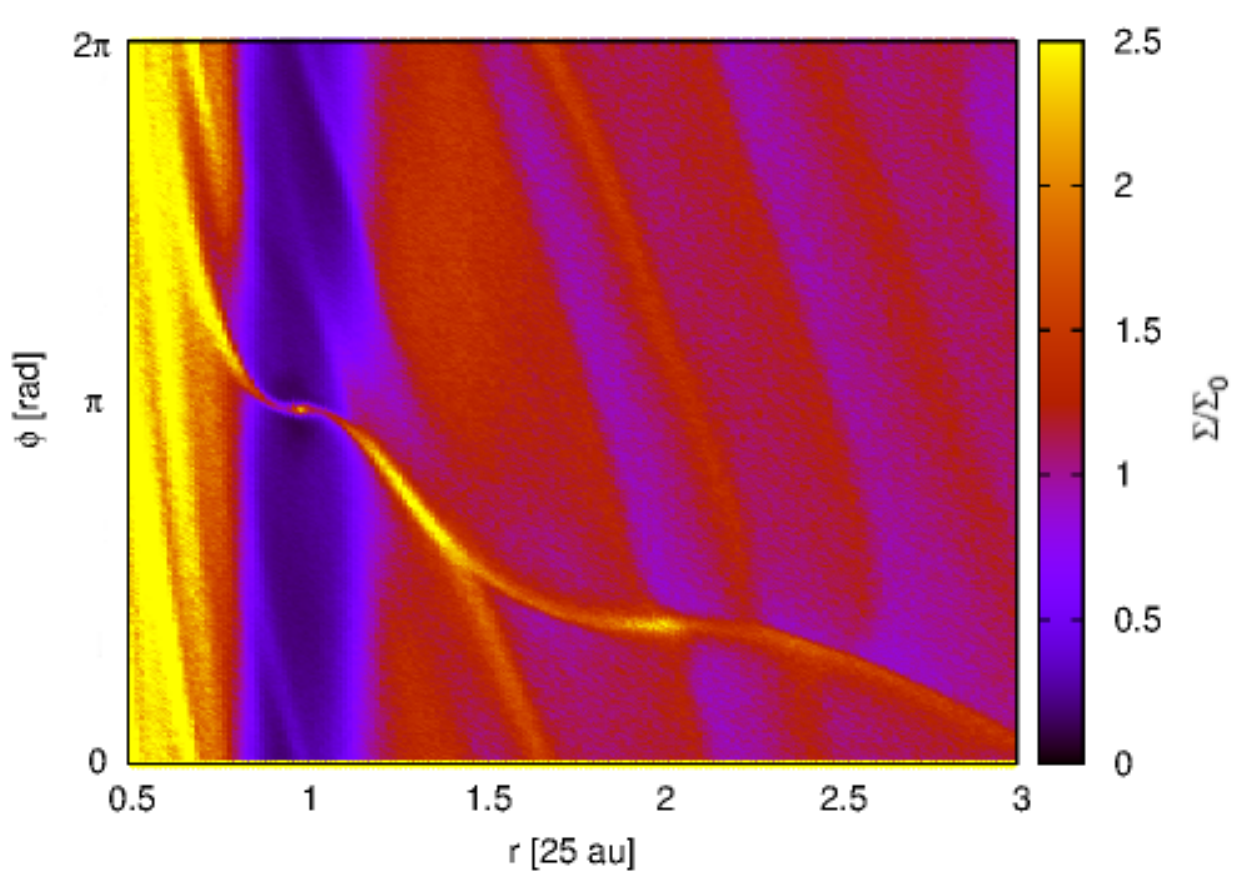}
        \includegraphics[width=.45\textwidth]{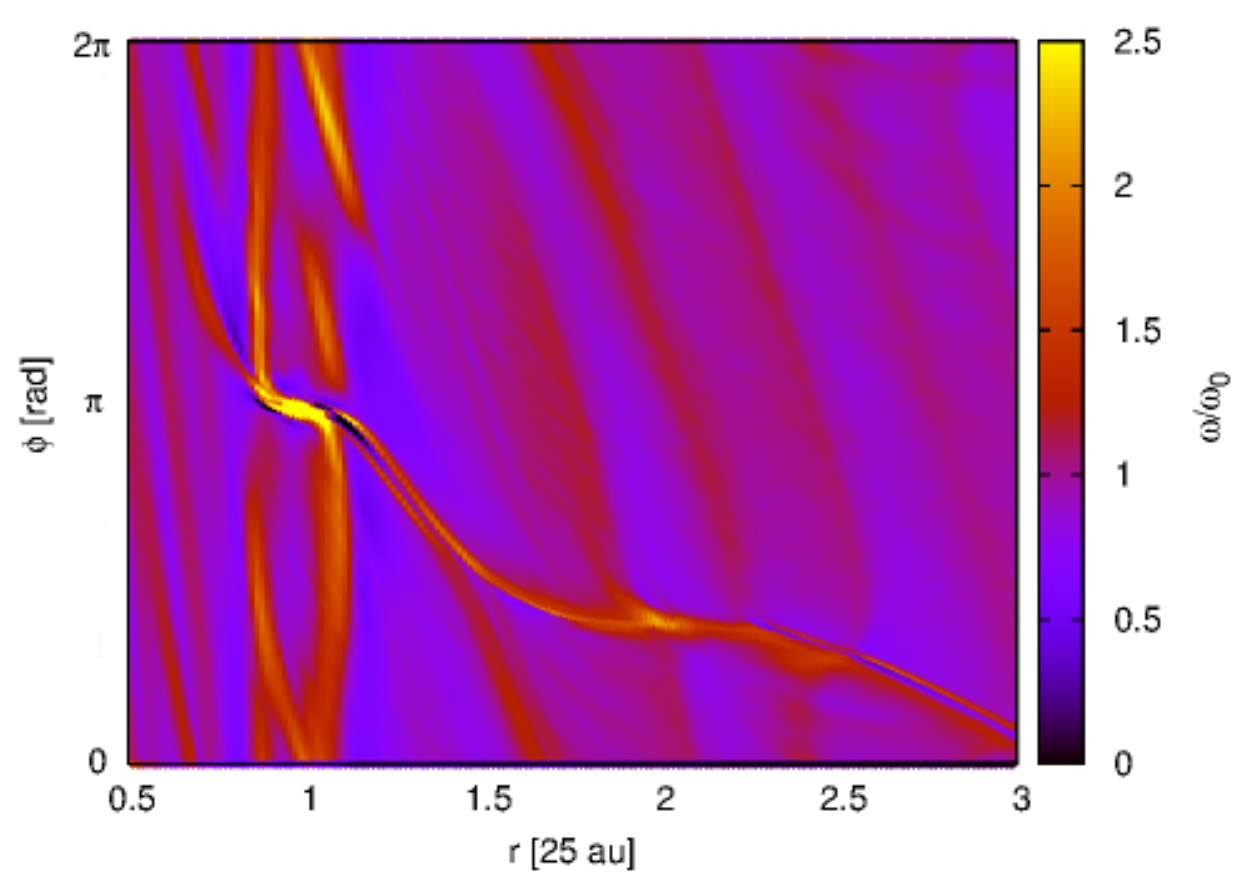}
        \caption{Gas surface density (top panel) and vorticity distribution
          (bottom panel) at the end of the intermediate mass cores
          simulation.~\label{Fig:Surf2D5}}
      \end{figure}

      %---------------------
      From the density profile of Figure~\ref{Fig:Surf10} (first panel) we can
      see that after $100$ orbits there is a visible gap opened by the inner
      planet, even if it is considerably shallower than the massive cores case,
      as for the vorticity profile (second panel).
      The outer planet still perturbs the co-orbital region of the inner one
      (Figure~\ref{Fig:Surf2D5}), but its magnitude is reduced and
      it has has barely modified the unperturbed surface density distribution
      at its location.

      Even though the influence by the outer planet is less dramatic compared
      to the more massive case, the reduction of the surface density profile's
      steepness hinders the development of vortices near the ring inner edge.

      %---------------------
      \subsubsection{Particle distribution}
\begin{figure}
        \centering
        \includegraphics[width=.38\textwidth]{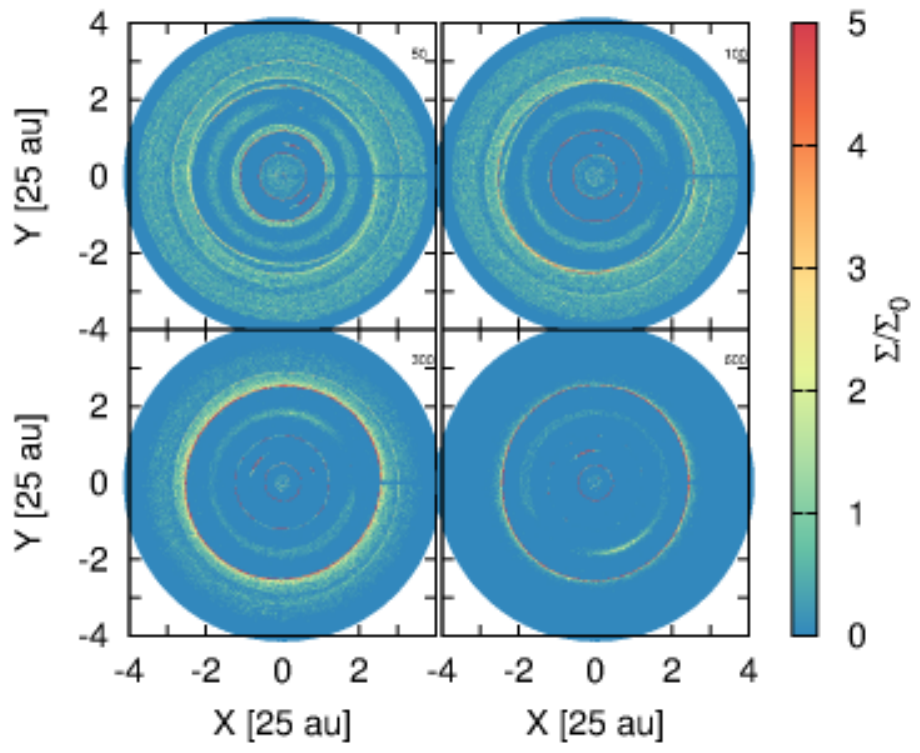}
        \includegraphics[width=.38\textwidth]{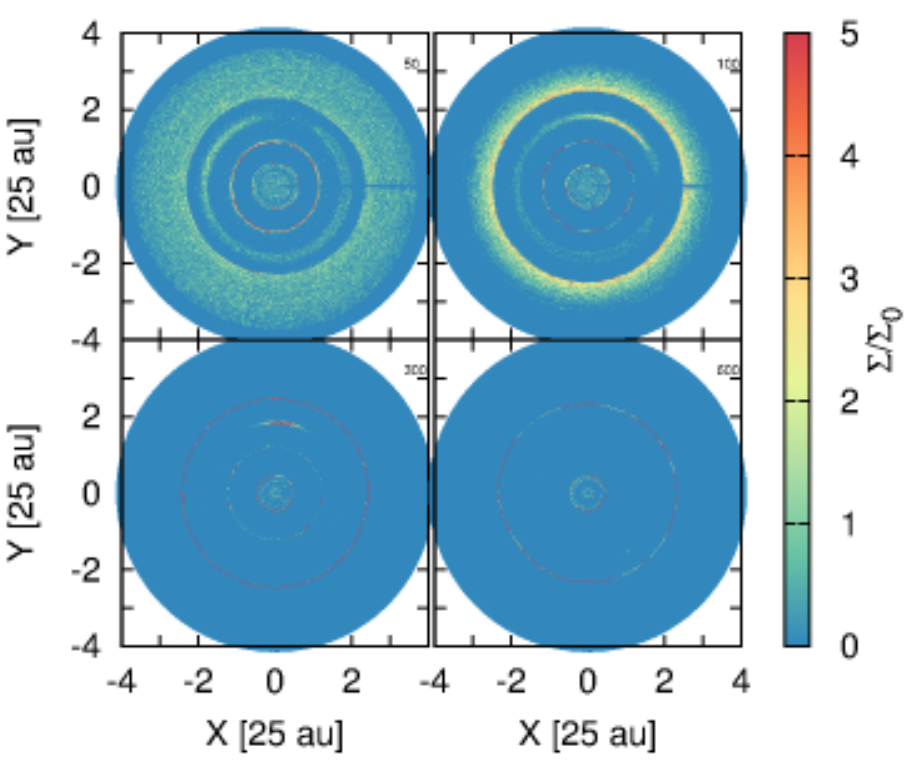}
        \includegraphics[width=.38\textwidth]{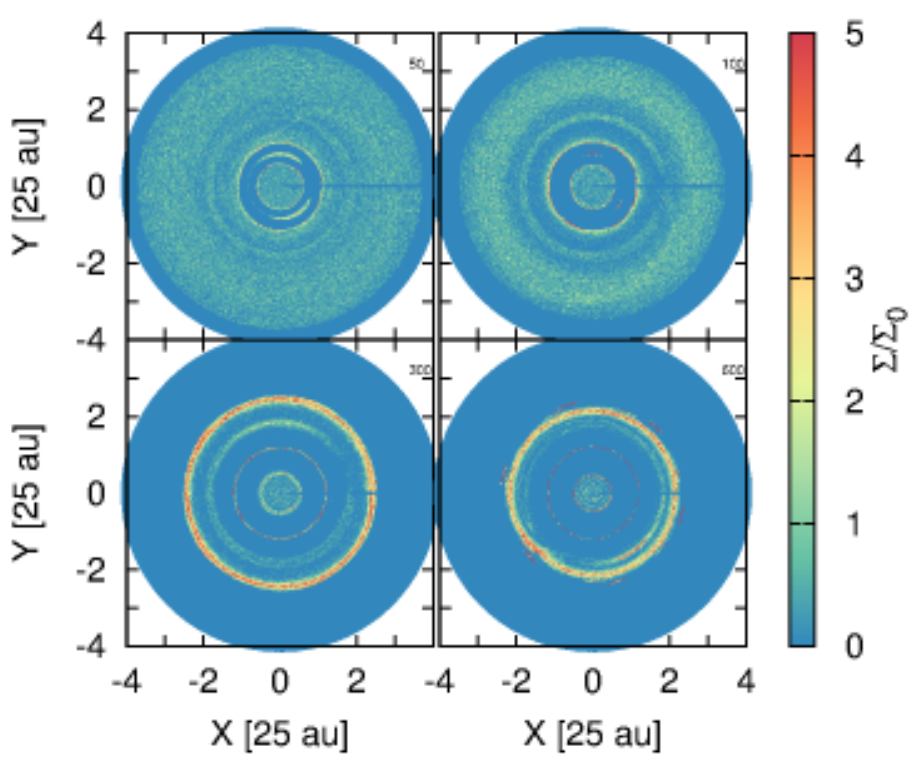}
        \includegraphics[width=.38\textwidth]{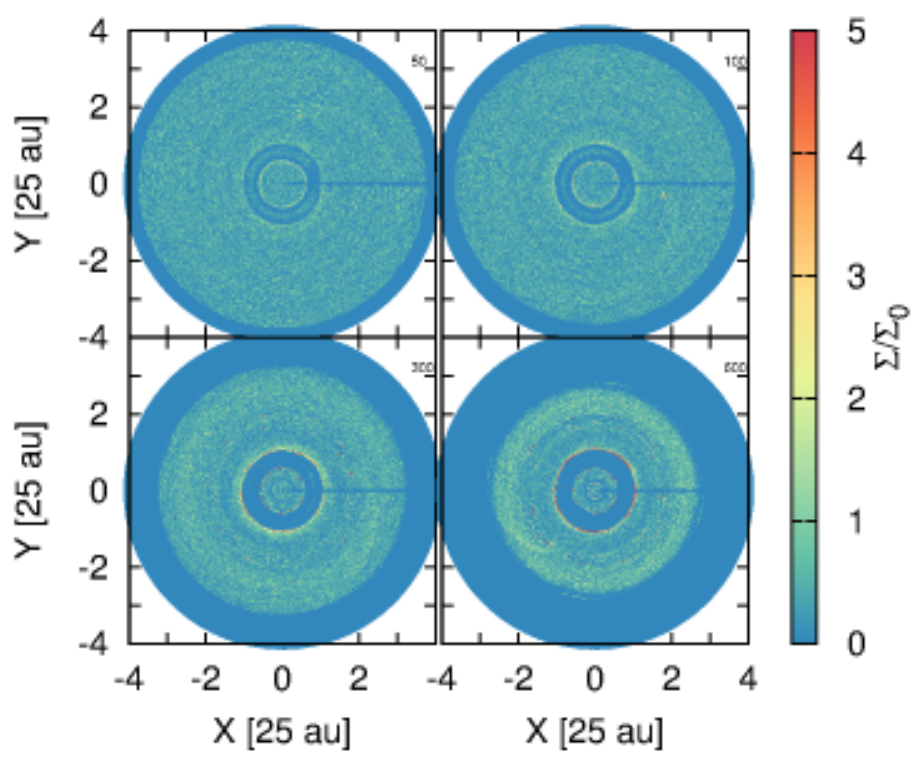}
        \caption{Dust normalised surface density distribution for the m (first
          panel), dm (second panel), cm (third panel), and mm-sized (fourth
          panel) particles disk and 2 equal mass $5\,M_\mathrm{th}$ cores at
          $r=1,2$ at 4 different times (50, 100, 300, 600 orbital times) for
          each case.~\label{Fig:Dust5m}}
      \end{figure}

      %---------------------
      The intermediate mass inner planet carves a gap in the dust disk
      after $50$ orbits for all the particles sizes (Figure~\ref{Fig:Dust5m}).
      Instead, the outer planet after $50$ orbits is able to clean its gap only
      for the most decoupled particles (dm and m), while it takes $~300$ orbits
      to clean a gap in the cm particles (third panel), and it has open only a
      partial gap in the mm size particles disk at the end of the simulation
      (fourth panel).
      The dm particles (second panel) have the highest migration speed in the
      outer disk, as expected from Figure~\ref{Fig:Vrad} and after $300$ orbits
      they are all distributed in narrow regions at gap edges and in co-orbital
      region with the outer planet.
      On the other hand, the cm particles (third panel) are more coupled to the
      gas and the outer planet has not yet carved a gap sufficient to confine
      them in the outer regions of the disk at the end of the simulation.
      Thus, we observe after $600$ orbits that they engulf the planet gap (third
      panel, last screen-shot).

\begin{figure}
        \centering
        \includegraphics[width=.45\textwidth]{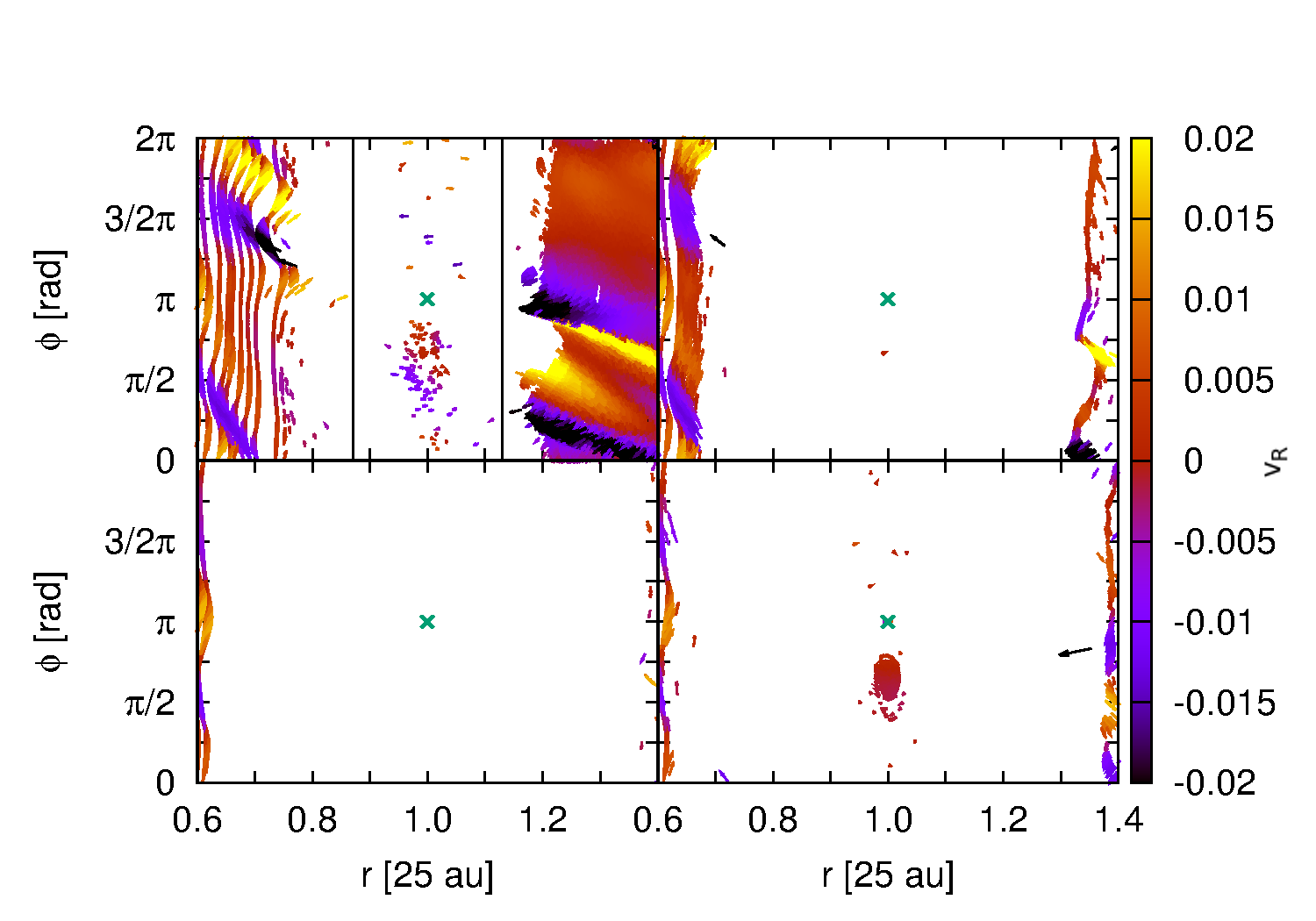}
        \caption{Particle distribution near the inner intermediate mass
          planetary core location for mm (top left), cm (top right), dm (bottom
          left) and m (bottom right) size particles at the end of the
          simulation.
          The velocity vectors of the particles respect to the planet are shown
          and the colour scale shows the relative radial
          velocity.~\label{Fig:Gap5m}}
      \end{figure}

      The dependence of the gap width respect to particle size is highlighted
      in Figure~\ref{Fig:Gap5m}, where we can see a strong difference between
      the mm particles gap, which follows the gas dynamic, remaining close to
      the location of the half-gap width ($x_\mathrm{s}$) over-plotted on the
      first panel, and the gap width of the other particles.

      The interaction between the two planet disrupts on the long term the
      particles clumping around the stable Lagrangian points, as in the more
      massive case.
      At the end of the simulation only the m and mm particles
      (Figure~\ref{Fig:Gap5m}) are still present in
      co-orbital region, preferable at the L5 point.

      The clumping of particles in few stable points at the ring location is
      still visible in the intermediate mass case, but only in the m, and dm
      cases (first and second panels of Figure~\ref{Fig:Dust5m}).

      Furthermore, some vortices are forming in the outer part of the particle
      disk of cm dust (third panel of Figure~\ref{Fig:Dust5m}).
      However, these are numerical artefacts because the disk outer
      edge is initialized with a sharp density profile cut which, on the long
      term, favours the development of vortices.

    %---------------------
    \subsection{Low mass core ($1 M_\mathrm{th}$)}
    %---------------------
    Finally, we explore the low mass core scenario in order to study a case
    where the particle ring between the two planets does not clump into few
    stable points and the gaps carved by planets remain narrow, such as in
    the observed HL Tau system.

      %---------------------
      \subsubsection{Gas distribution}
\begin{figure}
        \centering
        \includegraphics[width=.45\textwidth]{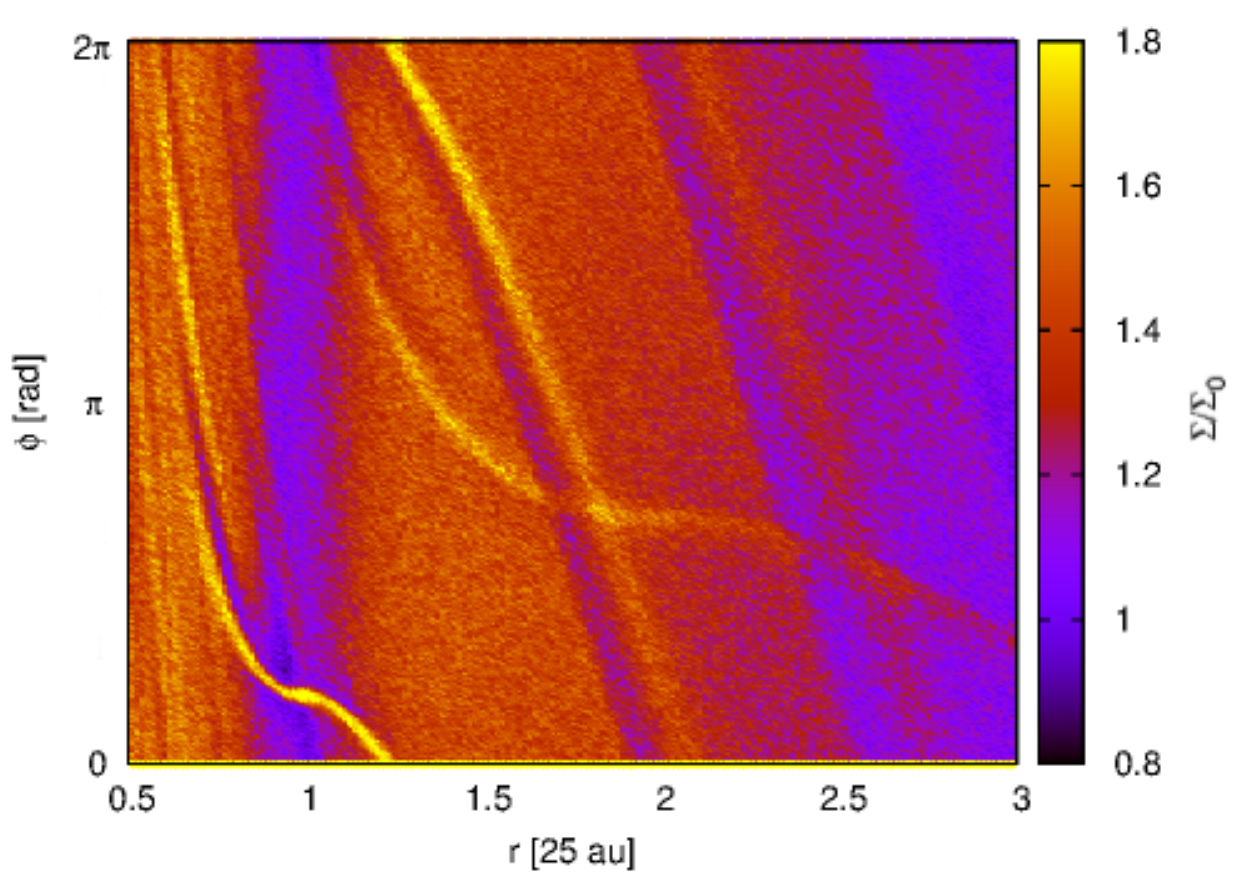}
        \includegraphics[width=.45\textwidth]{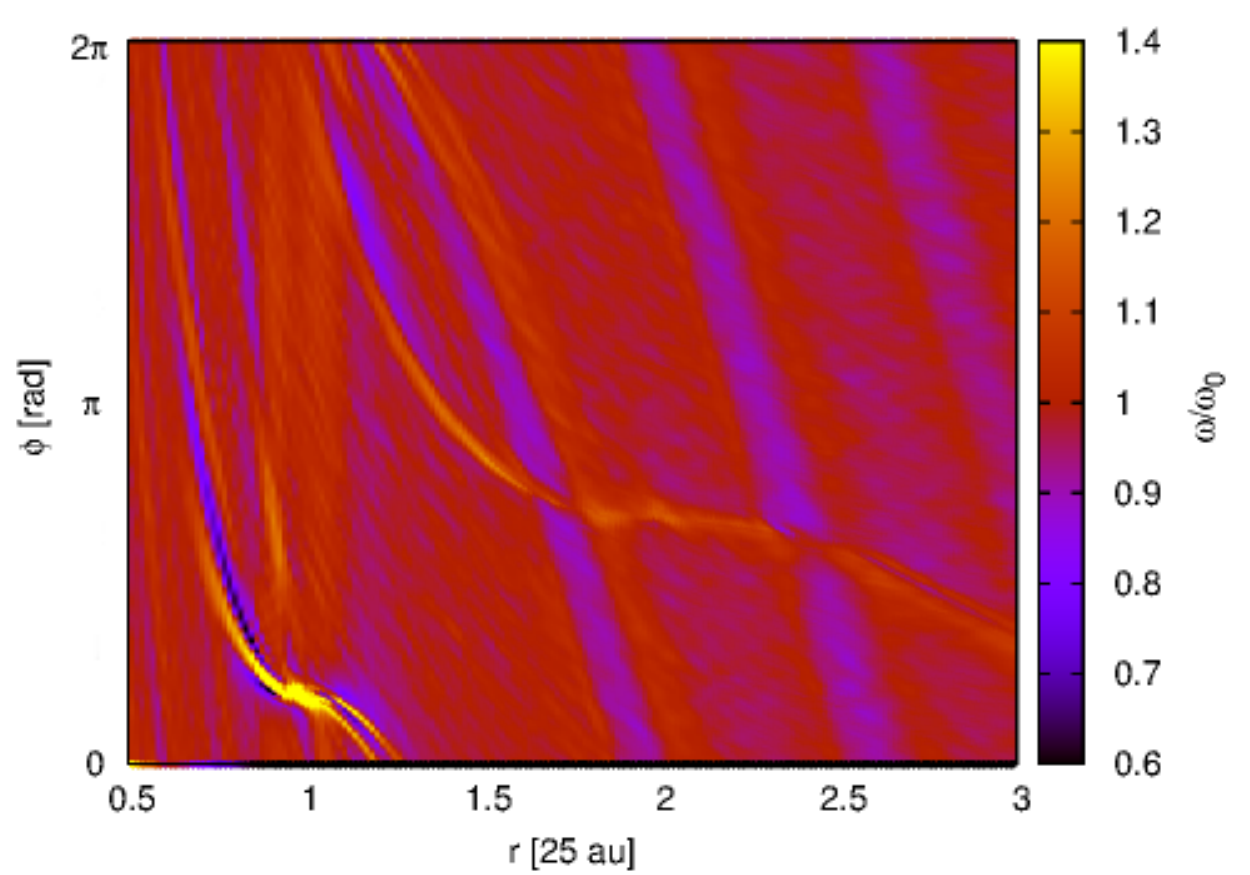}
        \caption{Gas surface density (top panel) and vorticity distribution
          (bottom panel) at the end of the simulation for the two low mass
          cores.~\label{Fig:Surf2D1}}
      \end{figure}

      %---------------------
      Figure~\ref{Fig:Surf10} (top panel) shows that both the inner and outer
      planet are not massive enough to clear a gap in the gaseous disk within
      the simulated time.
      From the 2D distribution of the surface density and vorticity
      (Figure~\ref{Fig:Surf2D1}) it is possible to see that the
      presence of the two planets changes slightly the unperturbed state of
      the gaseous disk (where the scale has been changed respect to the plots
      of the more massive cases in order to highlight the small differences).

      %---------------------
      \subsubsection{Particle distribution}
\begin{figure}
        \centering
        \includegraphics[width=.38\textwidth]{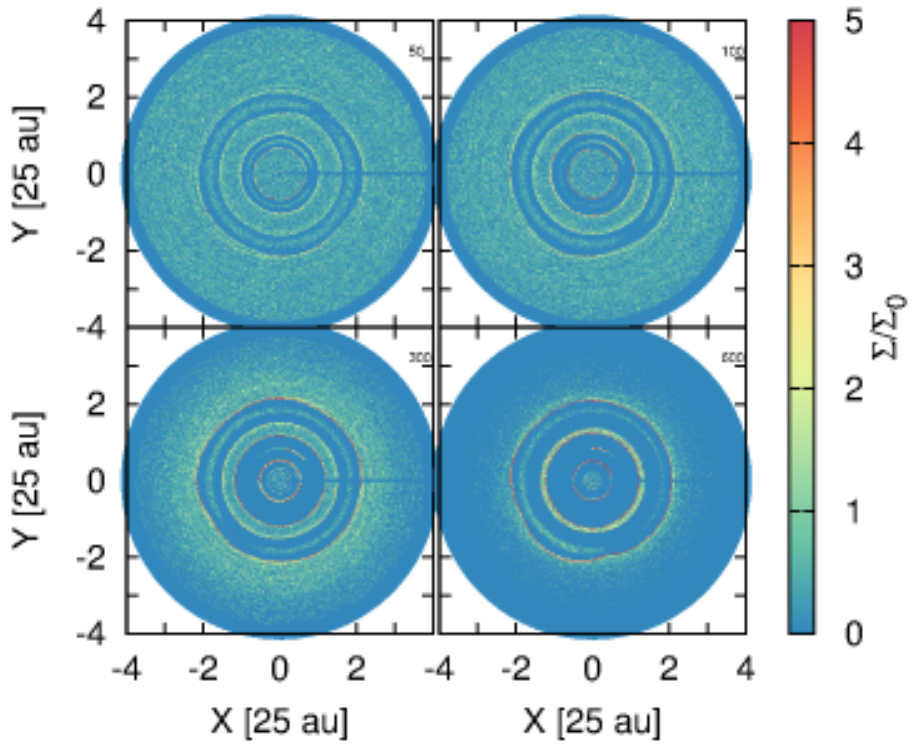}
        \includegraphics[width=.38\textwidth]{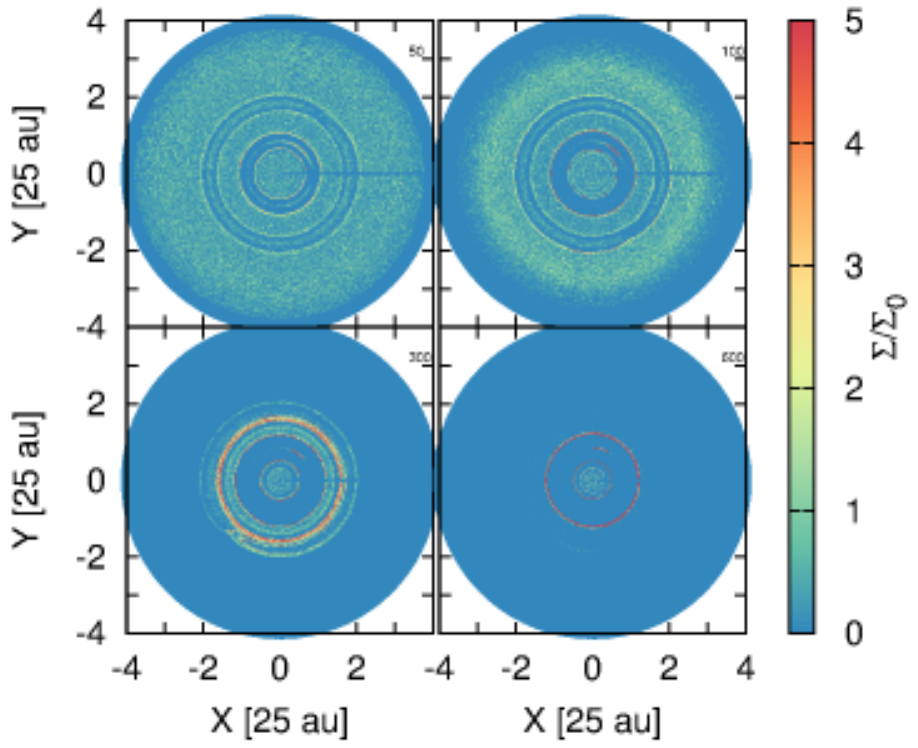}
        \includegraphics[width=.38\textwidth]{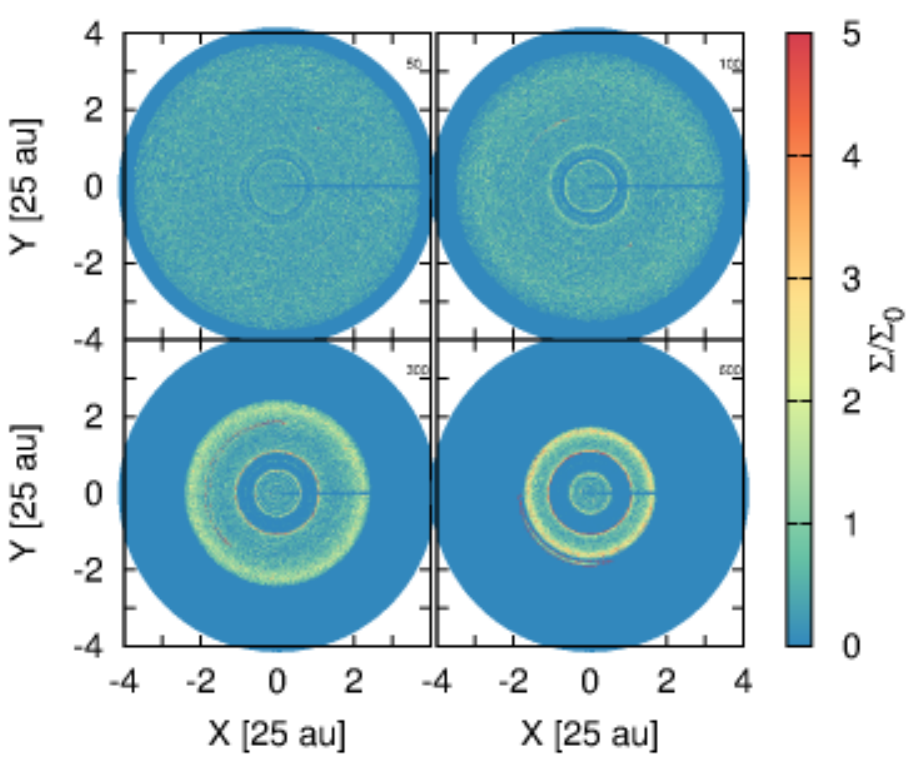}
        \includegraphics[width=.38\textwidth]{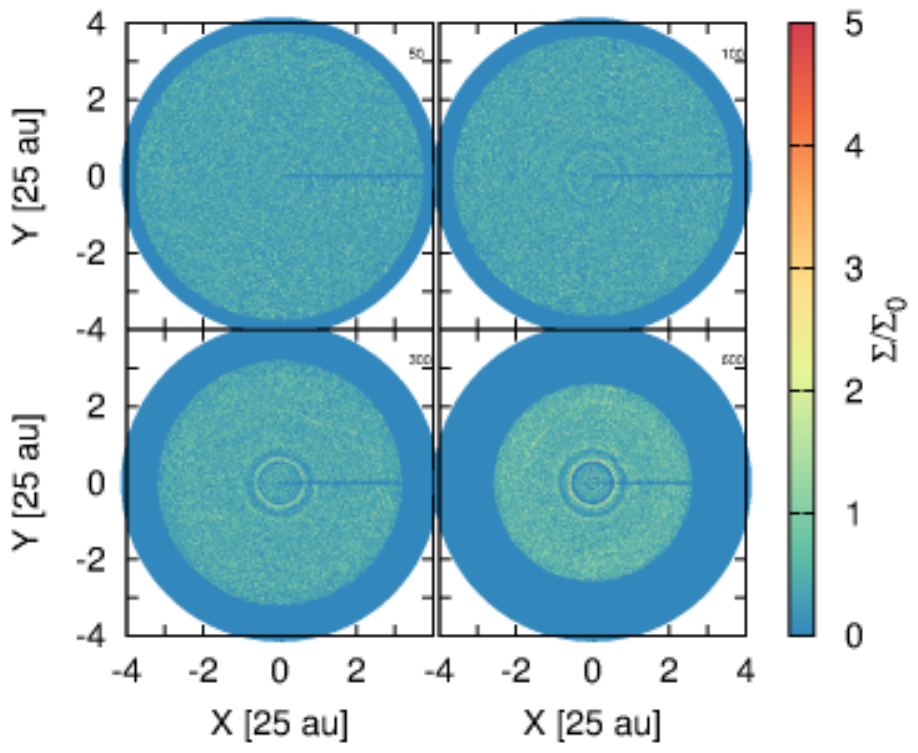}
        \caption{Dust normalised surface density distribution for the m (first
          panel), dm (second panel), cm (third panel), and mm-sized (fourth
          panel) particles disk and 2 equal mass $1\,M_\mathrm{th}$ cores at
          $r=1,2$ at 4 different times (50, 100, 300, 600 orbital times) for
          each case.~\label{Fig:Dust1m}}
      \end{figure}

      %---------------------
      Although the gas profile is not changed considerably due to the small mass
      of the planets, they are able to open up clear gaps in the dust disk in
      the first 50 orbits for m and dm size particles (Figure~\ref{Fig:Dust1m}),
      while it takes $100$ orbits for the cm size particles and at the end of
      the simulation it is still clearing the gap for the most coupled
      particles.
      As stated before, this process takes longer for the outer planet which is
      able to open a clear gap only at the end of the simulation for all but
      the mm size particles where only a partial gap is barely visible.

      A significant fraction of particles remains in the co-orbital region with
      both the inner and outer planet for the all period simulated.
      As found for the massive core case, their density is higher in the L4
      point.
      Also in this case, the dm size particles are the ones with the fastest
      dynamical evolution and it is possible to see in Figure~\ref{Fig:Dust1m}
      (third panel) that after 300 orbits the outer disk engulfs the outer
      planet co-orbital region, which is unable to filtrate effectively those
      particles.

\begin{figure}
        \centering
        \includegraphics[width=.45\textwidth]{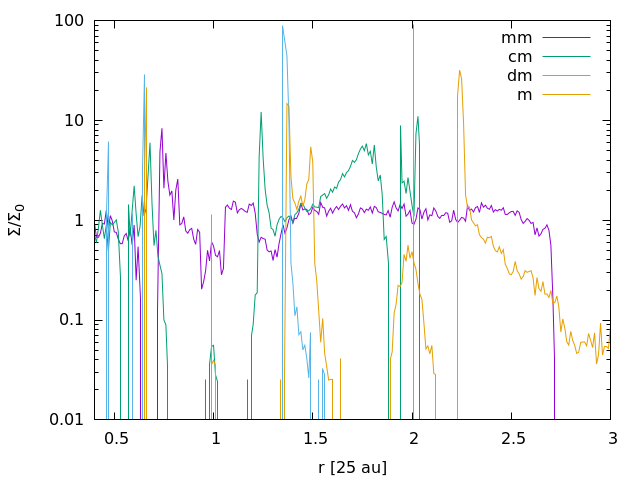}
        \includegraphics[width=.45\textwidth]{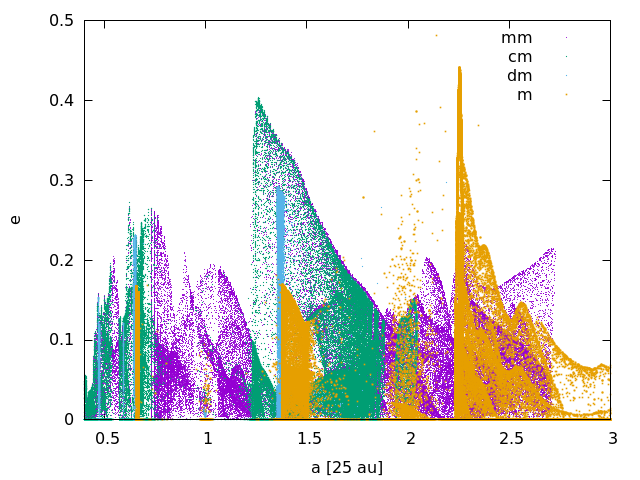}
        \caption{Particle surface density (top panel) and eccentricity profile
          (bottom panel) at the end of the simulation for the different particle
          species.\label{Fig:SurfPart1}}
      \end{figure}

      In this case, the ring between the two planets does not clump in a small
      number of stable points, but it remains wide for the all simulation,
      though it shrinks with time and its width depends on the particle size
      (Figure~\ref{Fig:SurfPart1}~-top panel).

      From Figure~\ref{Fig:Dust1m} (first panel) is it possible to see also the
      formation of ripples just outside the outer planet location.
      This effect can be recalled from the final eccentricity distribution
      of the m size particles (Figure~\ref{Fig:SurfPart1}~-~bottom panel).
      This behaviour is due to the eccentricity excitation of particles that
      passes close to the planet location.
      For less coupled particles, the interaction with the gas takes several
      orbital time-steps in order to smooth out the eccentricity, thus these
      typical structures form.
      This effect is only visible in the low mass cores simulation since the
      particle gap is narrower, thus the particles get closer to the planet
      location and the excitation of their eccentricity is higher.

\begin{figure}
        \centering
        \includegraphics[width=.45\textwidth]{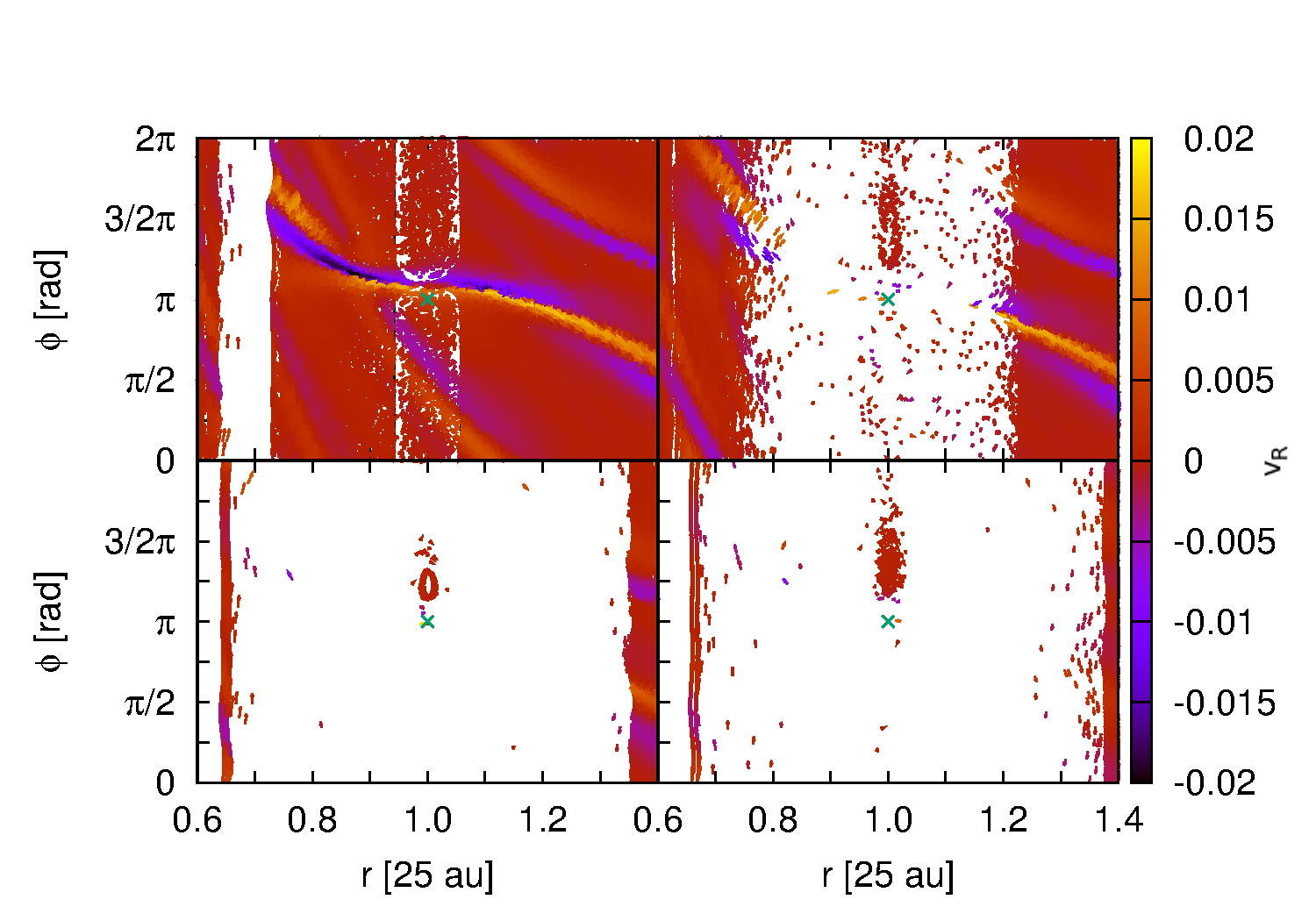}
        \caption{Particle distribution near the inner low mass planetary
          core location for mm (top left), cm (top right), dm (bottom left) and
          m (bottom right) size particles at the end of the simulation.
          The velocity vectors of the particles respect to the planet are shown
          and the colour scale shows the relative radial
          velocity.~\label{Fig:Gap1m}}
      \end{figure}

      In Figure~\ref{Fig:Gap1m} we focused on the gap structure close to the
      inner planet location for the different particle sizes at the end of the
      simulation.
      We overplot also the minimum gap half-width $x_\mathrm{s}$, in order to
      test whether this condition is met for the most coupled particles which
      follow the gas dynamics.
      From the distribution of mm size particles we can see that the gap is
      opened exactly at the location of the minimum gap half-width, and there
      is still a lot of material in the horseshoe region.
      The gap is considerably wider for the cm size particles

\begin{figure}
        \centering
        \includegraphics[width=.45\textwidth]{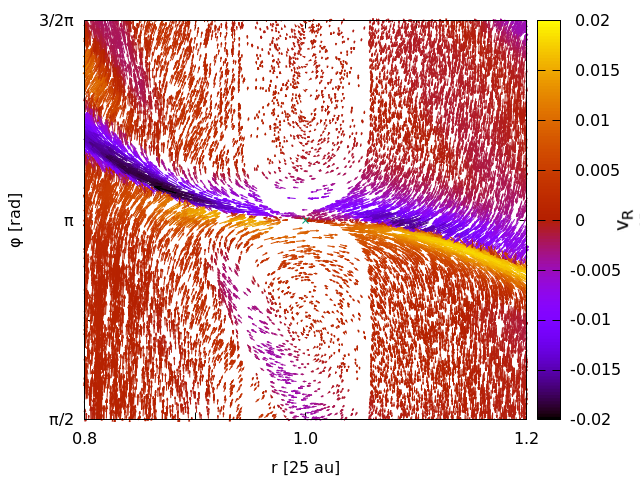}
        \caption{Mass transfer through the planet position of mm-size particles
          for the 1 thermal mass planet, which is unable to effectively filtrate
          them. The particles are plotted together with their velocity vectors
          and the colorscale indicates their radial
          velocities.\label{Fig:Masstrans}}
      \end{figure}

      Finally, in Figure~\ref{Fig:Masstrans} we highlighted the constant mass
      transfer that take place through the inner planet location for the most
      coupled particles which are not effectively filtered by the less massive
      planet.
      We plot the radial velocity of the particles with colour a scale in
      order to emphasise the flow in both directions.

  %%%%%%%%%%%%%%%%%%%%%%
  \section{Discussion}\label{sec:disc}
  %%%%%%%%%%%%%%%%%%%%%%

    %---------------------
    \subsection{Path to a second generation of planets}\label{par:plan}
    %---------------------
    The possibility to create the conditions for a second generation
    of planets by a massive core has already been studied in the past.
    However, the combined action of a multi-planetary system can achieve the
    same goal from less massive first generation planets.
    As we have seen in Sec.~\ref{sec:res10}, two
    $10\,M_\mathrm{th}=0.7\,M_\mathrm{Jup}$ can trigger vortex formation which
    are more prominent and live longer compared to the single mass case.
    Nevertheless, these vortices appear like a transient effect that develop
    after more than $80$ orbits of the inner planet and last until $120$ orbits,
    thus it is difficult to relate this behavior to an initial condition effect
    and it does not depend on the timescale over which the planet are grown in
    the simulation.
    Furthermore, vortex formation depends on viscosity and it can persist for
    longer times in low viscosity disks.
    Although the main dust collection driver is the ring generated by the
    combined action of the two planets, vortices can create some characteristic
    observable features, enhancing locally the dust surface density.
    Moreover, for both the massive and intermediate mass cores, the particle
    ring clumps into few symmetric points close to the 5:3 MMR which are stable
    for several hundred orbits.
    Taking a dust-to-gas mass ratio of $0.01$ we found that the mass collected
    in those stable points can reach several Earth masses.
    However, we point out that this strong mass clumping might be reduced by
    the introduction of particle diffusion due to disk turbulence.

    %---------------------
    \subsection{Comparison with Previous Work}\label{par:prev}
    %---------------------
    This is one of the first studies on the dust evolution and filtration in a
    multiple planet system thus there are no direct comparison with similar
    setups.
    However, recently~\cite{Zhu2014} have performed an extensive analysis of
    the dust filtration by a single planet in a 2D and 3D disk, from which we
    have taken some ideas and it is the natural test comparison for the
    different outcomes of our analysis.

    One first interesting comparison between the two results is the
    possibility to form vortices at the gap edges, taking into account that in
    their scenario vortex development was favoured by the choice of a
    non-viscous disk.
    We found that, even if for single planet the vortices are hindered, the
    presence of an additional planet can enhance the density at the ring
    location and promote the development of vortices.

    Moreover, we regain the `ripple' formation as in~\citet{Zhu2014} for
    decoupled particles close to the planet location.

    \cite{Zhu2014} found also a direct proportionality between the gap width and
    the planet mass, where a $9 M_\mathrm{Jup}$ induced a vortex at the gap
    edge at a distance more than twice the planet semi-major axis.
    Although we have not modelled such high mass planets we have obtained a
    similar outcome for our parameter space.

    In our simulation we do not find the presence of strong MMR which aid the
    gap clearing since we do not model particles with stopping times greater
    than $\sim$ 5, thus the coupling with the gas disrupt the MMR~.
    They became important only when the gas surface density is highly depleted,
    such as in the ring between planets for the high mass cases, where a 5:3
    MMR with the inner planet is found to be a stable location.

    Furthermore, as observed in~\cite{Ayliffe2012} there is a strong
    correlation between particle size and particle gap, where the most coupled
    particles reach regions closer to the planet location, and are potentially
    accreted by the planet or they migrate in the inner disk, while less
    coupled particles are effectively filtrated by the planet.

    %---------------------
    \subsection{Comparison with Observations}\label{par:obs}
    %---------------------
    Pre-transitional disks are defined observationally as disks with gaps.
    These features are observed in many cases in the sub-mm dust emission and
    there is no evidence that the gaseous emission follows the same pattern.
    The observation of (pre-)~transitional disks highlights different physical
    behaviours that need to be explained.
    A major problem is the coexistence of a significant accretion rate onto the
    star (up to the same order as common T Tauri Stars~-~CTTS) with dust cleared
    zones, and the absence of near infra-red (NIR) emission.
    One of the most plausible explanation to solve this issue is the presence of
    multiple giant planets that can create a common gap and thus enhance the
    accretion rate across them exchanging torque with the disk, while depleting
    the dust component, through a filtration mechanism that, together with dust
    growth, can explain the absence of a strong NIR emission in
    (pre-)~transitional disks compared to full disks \citep{Zhu2012}.
    For a full review of the topic see~\cite{Espaillat2014}

    In the HL Tau system, although several rings have been observed in its dust
    emission, it has still a very high accretion rate onto the star
    $\dot{M} = 2.13*10^{-6}M_\odot/yr$ \citep{Robitaille2007}.
    This is a prove that, at least in this system, the rings observed in
    the dust emission are not related to rings in the gas distribution.

\begin{figure}
      \centering
      \includegraphics[width=.45\textwidth]{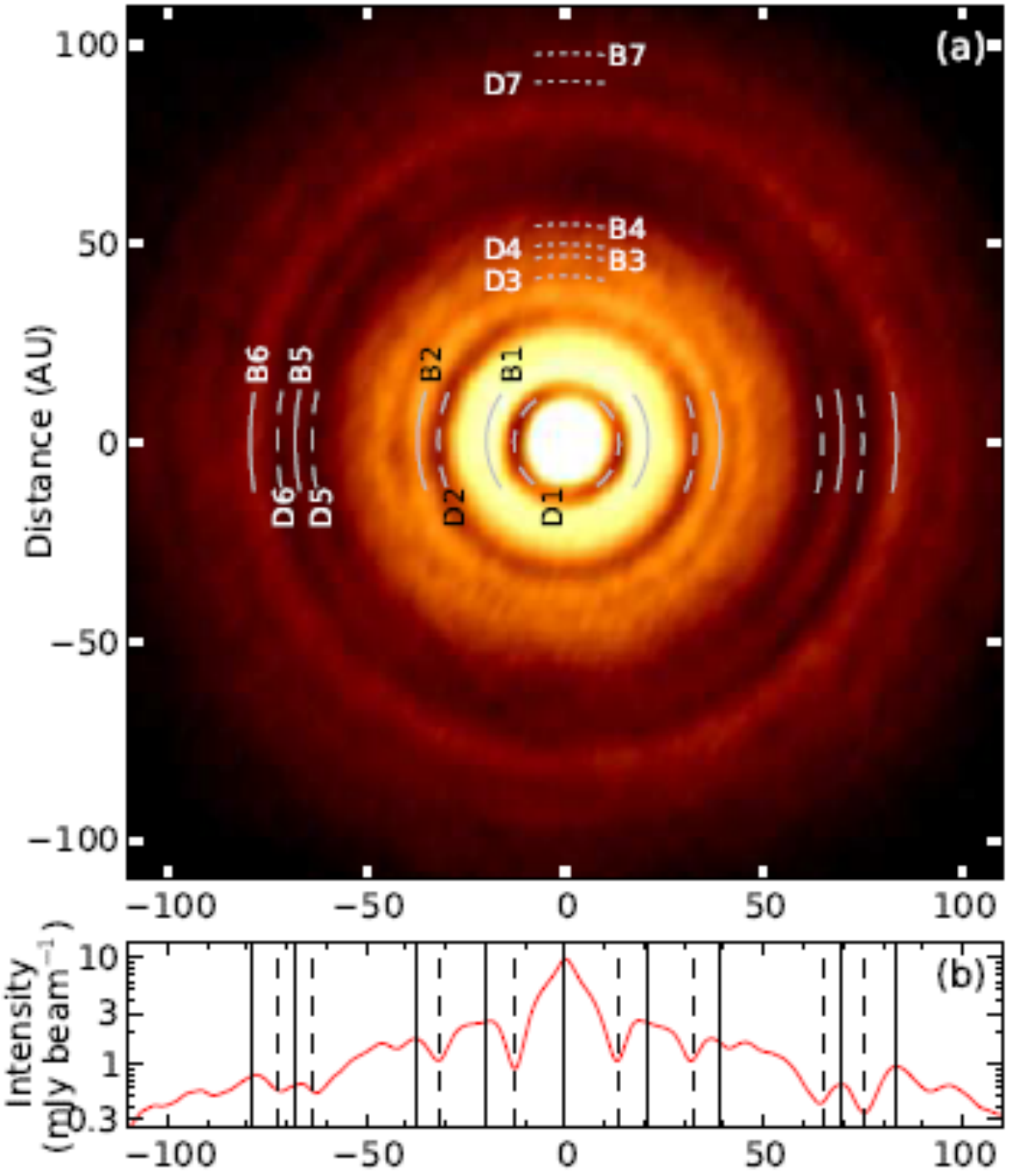}
      \caption{Top panel: deprojected image from the mm continuum of HL Tau.
        Bottom panel: cross-cuts at PA=$138^\circ$ through the peak of the mm
        continuum of HL Tau~\cite{Partnership2014}.~\label{Fig:HLTau2}}
    \end{figure}

\begin{figure}
      \centering
      \includegraphics[width=.45\textwidth]{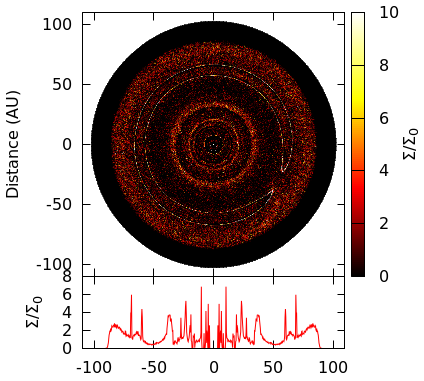}
      \caption{Top panel: final mm-dust surface density distribution for a inner
        low mass core and an outer intermediate mass core.
        Bottom panel: relative surface density distribution.~\label{Fig:HLTau3}}
    \end{figure}

    Since we found that a wide ring is observed in our simulations only for
    the small mass case and a clear gap is visible for the outer planet only
    in the intermediate and massive core cases, we choose to run a different
    model with an inner small mass core and an outer intermediate mass core
    in order to compare the outcome of our simulations with the HL Tau system.
    We rescaled the system and run a different simulation in order to compare it
    to the real one, placing the inner planet at $32.3\,\mbox{au}$ corresponding
    to the D2 gap (see Figure~\ref{Fig:HLTau2}~-~top panel) and the outer one at
    $64.6\,\mbox{au}$, keeping the 2:1 ratio, which is close to the B5 location.
    Comparing the 2D surface density distribution at the end of the simulation
    with the deprojected image in the continuum emission and their slices
    (Figures~\ref{Fig:HLTau2}-\ref{Fig:HLTau3}) we can outline several shared
    features and differences.
    The gap created by the inner planet has a very similar configuration as the
    one observed.
    On the other hand there are no clear visible features inside its orbit in
    the observed imaged while a variety of inner structures are visible in the
    output of the simulation.
    These features are mainly due to the inner wave generated by the planet.
    The strong gap that is visible close to the star is instead not physical
    and it is related to the inner boundary condition.
    In the outer part of the disk, several differences can be oulined.
    The major one is the high surface density in the horseshoe region, which is
    related to our choice of an initial flat profile for the particle
    distribution.
    Although this approximation was chosen to extend the simulated time
    preventing a fast depletion of material from the outer disk, it also
    favoured the dust trapping by the outer planet.
    Moreover, the particle ring is more depleted than in the observed image.
    Thus, we expect that the planetary mass responsible for the observed
    outer gap should be slightly smaller than the one adopted in this
    simulation.
    A final remark is the strong depletion of dust particles just inside the
    outer planet location due to its dust filtration mechanism, which is clear
    from the bottom panel of Figure~\ref{Fig:HLTau2}.
    Due to this effect, in a multiplanetary system, a planet is not necessary
    located where the gap is deeper but it could be at the rim of the gap
    preventing the particles of a certain size to cross its location.

    However, a part from these differences, due to our initial choice of
    the parameter space, the structures obtained from the simulations are
    similar to what is observed.

    %---------------------
    \subsection{Dependence on the disk surface density}\label{par:sta}
    %---------------------
    The dynamical evolution of dust particles is closely linked to their
    stopping time, which is directly related to the disk surface density
    through eq.~\ref{eq:stop}, for the Epstein regime.
    Thus, if we decrease the disk surface density by a factor $10$ in order to
    stabilize the disk in the isothermal case (see Appendix~\ref{sec::stability}
    for a study of the disk stability with a more realistic equation of state),
    we need to lower by the same factor the particle size to reobtain the same
    particle dynamics.
    On the other hand, if we want to keep the particle's size fixed, in
    order to compare our results with the ALMA continuum images, we need
    to decrease the planetary mass, since the gap width depends on the
    particle stopping time.
    We tried a different choice of the parameter space to obtain a similar
    output with a much smaller disk mass and planetary mass cores.
    \begin{figure}
      \centering
      \includegraphics[width=.45\textwidth]{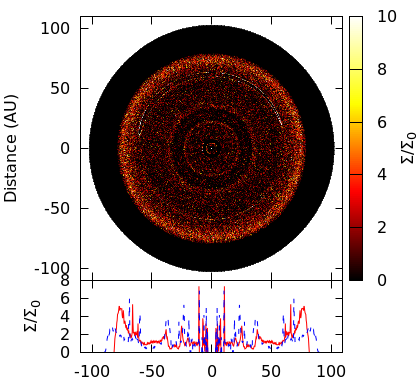}
      \caption{Top panel: dust distribution for the mm-sized particles disk and
      2 planetary cores of $\sim 10$ and $\sim 20\,M_\oplus$ after 250 orbits
      of the inner planet. Bottom panel: Relative surface density distribution
      (red solid curve), where the distribution of the higher disk mass case (from
      Figure~\ref{Fig:HLTau3} has been overplotted (dashed blue curve).
      ~\label{Fig:HLTau4}}
    \end{figure}
    We report in Fig.~\ref{Fig:HLTau4} a run with a disk mass of
    $1/10$ of the test case.
    The width of the gap created by two planets is similar, as outlined by
    the bottom panel.
    However, in this case the planetary mass adopted to open such narrow gaps
    in the particle disk are much lower: $\sim 10$ and
    $\sim 20\,M_\oplus$ for the inner and ouer planet, respectively.
    These lower values of the planetary masses prove that the ability of planets
    to open gaps in the dust disk is widely applicable, and increase the
    likelihood of the planetary origin through core accretion in this young
    system at large radii ($\sim 60\,\mbox{au}$).
    It will be crucial to define better the disk mass in order to constrain the
    particle dynamics and the planetary mass growing inside their birth disk.

  %%%%%%%%%%%%%%%%%%%%%%
  \section{Conclusion}\label{sec:conc}
  %%%%%%%%%%%%%%%%%%%%%%
  We have implemented a population of dust particles into the 2D hydro code
  \textsc{fargo}~\citep{Masset2000} in order to study the coupled dynamic of
  dust and gas.
  The dust is modelled through Lagrangian particles, which permit us to cover
  the evolution of both small dust grains and large bodies within the same
  framework.
  We have studied in particular the dust filtration in a multi-planetary system
  to obtain some observable features that can be used to interpret
  the observations made by modern infrared facilities like ALMA~.
  From the analysis of our simulations we have found that the outer planet
  \begin{itemize}
    \item affects the co-orbital region of the inner one exciting the particles
      in the Lagrangian points (L4 and L5), which are effectively removed in the
      majority of the cases,
    \item increases the surface density in the region between them, creating a
      particle ring which can clump in a small number of symmetric points,
      collecting a mass up to several Earth masses,
    \item promotes the development of vortices at the ring inner edge,
      increasing the steepness of the surface density profile.
  \end{itemize}
  Moreover, when the planets are not massive enough to create a narrow
  particle ring between the planets, its width depends on the particle size.
  This could be a potentially observable feature that can link the ring
  formation with the presence of planets.
  Furthermore, we confirmed previous results regarding the particle gap, which
  develops much more quickly than the gaseous one, and is wider for higher
  mass planets and more decoupled particles.

  The features observed in the HL Tau system can be explained through the
  presence of several massive cores, or lower mass cores depending on the
  adopted surface density, that shape the dust disk.
  We have obtained that the inner planet(s) should be on the order of
  $1\,M_\mathrm{th}=0.07\,M_\mathrm{Jup}$, in order to open a
  small gap in the dust disk while keeping a wide particle ring.
  The outer one(s) should have a mass on the order of
  $5\,M_\mathrm{th}=0.35\,M_\mathrm{Jup}$ in order to open a visible gap.
  These values are in agreement with those found
  by~\cite{Kanagawa2015,DiPierro2015,Dong2015}.
  We point out that, decreasing the disk surface density by a factor 10
  reduces the required planetary mass to open the observed gaps to a value
  of $10\,M_\oplus$ and $20\,M_\oplus$ respectively.
  These reduced values render the planet formation through core accretion
  more reliable in the young HL Tau system.
  Although the particle gaps observed are prominent, the expected gaseous gaps
  would be barely visible.

  The limitations of this work are the lack of particle back-reaction on the
  gas, self-gravity of the disk, and particle diffusion.
  Furthermore we have not model accretion of particles onto the planet and
  planet migration.
  These approximation were chosen in order to study the global evolution of
  particle distribution with different stopping times and different planet
  masses, without increasing excessively the computation time.
  Although the disk is very massive, the asymmetric features typical of a
  gravitationally unstable disk are not observed in the continuum mm
  observations that should correctly describe the gas flow, thus we do not
  expect the real system to be subject to strong perturbations due to its
  self-gravity.
  The particle-back reaction plays an important factor when studying the
  evolution of particle clumps, but it is not expected to change significantly
  the global dust distribution.
  However, particle diffusion could have an important role both
  in reducing the dust migration and preventing strong clumping of particles.
  We are planning in future works to relax these approximations, running more
  accurate simulations and testing the contribution of the individual physical
  process on the final dust filtration and distribution.
  Moreover, we have limited our analysis to the peculiar case of equal mass
  planets on a fixed orbit, and changing each one of these conditions can
  result in a rather different outcome.
  The possibility to evolve the simulations further in time has been considered
  since the outer planets where not able to open up a clear gap for the
  less massive cases.
  However, in any case they are able to open only a very shallow gap, so the
  possible influence on the subsequent evolution of the particles close to their
  position is not expected to be significant.
  Furthermore, since we have modeled the system for $\sim 10^5\,\mbox{yr}$
  around a young star ($\sim 10^6\,\mbox{yr}$), and we need to form the
  planetary cores in the first place, it does not seem unrealistic to observe a
  planet at $\sim 60\,\mbox{au}$ still in the gap clearing phase.

  %%%%%%%%%%%%%%%%%%%%%%
  %Acknowledgments
  %%%%%%%%%%%%%%%%%%%%%%
  \begin{acknowledgements}
    We thank an anonymous referee for his useful comments and suggestions.
    G. Picogna acknowledges the support through the German Research Foundation
    (DFG) grant KL 650/21 within the collaborative research program ''The first
    10 Million Years of the Solar System''.
    Some simulations were performed on the bwGRiD cluster in T\"ubingen, which
    is funded by the Ministry for Education and Research of Germany and the
    Ministry for Science, Research and Arts of the state Baden-W\"urttemberg,
    and the cluster of the Forschergruppe FOR 759 ''The Formation of Planets:
    The Critical First Growth Phase'' funded by the DFG.
  \end{acknowledgements}

  %%%%%%%%%%%%%%%%%%%%%%
  %Bibliography
  %%%%%%%%%%%%%%%%%%%%%%
  \bibliographystyle{aa}
  \bibliography{biblio}

\begin{thebibliography}{39}
\expandafter\ifx\csname natexlab\endcsname\relax\def\natexlab#1{#1}\fi

\bibitem[{{ALMA Partnership} {et~al.}(2015){ALMA Partnership}, {Brogan},
  {P{\'e}rez}, {Hunter}, {Dent}, {Hales}, {Hills}, {Corder}, {Fomalont},
  {Vlahakis}, {Asaki}, {Barkats}, {Hirota}, {Hodge}, {Impellizzeri}, {Kneissl},
  {Liuzzo}, {Lucas}, {Marcelino}, {Matsushita}, {Nakanishi}, {Phillips},
  {Richards}, {Toledo}, {Aladro}, {Broguiere}, {Cortes}, {Cortes}, {Espada},
  {Galarza}, {Garcia-Appadoo}, {Guzman-Ramirez}, {Humphreys}, {Jung}, {Kameno},
  {Laing}, {Leon}, {Marconi}, {Mignano}, {Nikolic}, {Nyman}, {Radiszcz},
  {Remijan}, {Rod{\'o}n}, {Sawada}, {Takahashi}, {Tilanus}, {Vila Vilaro},
  {Watson}, {Wiklind}, {Akiyama}, {Chapillon}, {de Gregorio-Monsalvo}, {Di
  Francesco}, {Gueth}, {Kawamura}, {Lee}, {Nguyen Luong}, {Mangum}, {Pietu},
  {Sanhueza}, {Saigo}, {Takakuwa}, {Ubach}, {van Kempen}, {Wootten},
  {Castro-Carrizo}, {Francke}, {Gallardo}, {Garcia}, {Gonzalez}, {Hill},
  {Kaminski}, {Kurono}, {Liu}, {Lopez}, {Morales}, {Plarre}, {Schieven},
  {Testi}, {Videla}, {Villard}, {Andreani}, {Hibbard}, \&
  {Tatematsu}}]{Partnership2014}
{ALMA Partnership}, {Brogan}, C.~L., {P{\'e}rez}, L.~M., {et~al.} 2015, \apjl,
  808, L3

\bibitem[{{Armitage}(2010)}]{Armitage2010}
{Armitage}, P.~J. 2010, {Astrophysics of Planet Formation} (Cambridge
  University Press)

\bibitem[{{Ayliffe} {et~al.}(2012){Ayliffe}, {Laibe}, {Price}, \&
  {Bate}}]{Ayliffe2012}
{Ayliffe}, B.~A., {Laibe}, G., {Price}, D.~J., \& {Bate}, M.~R. 2012, Monthly
  Notices of the Royal Astronomical Society, 423, 1450

\bibitem[{{Bai} \& {Stone}(2010)}]{Bai2010a}
{Bai}, X.-N. \& {Stone}, J.~M. 2010, The Astrophysical Journal Supplement
  Series, 190, 297

\bibitem[{Baruteau \& Masset(2008)}]{Baruteau2008}
Baruteau, C. \& Masset, F.~S. 2008, The Astronomical Journal, 672, 1054

\bibitem[{{Crida} {et~al.}(2006){Crida}, {Morbidelli}, \& {Masset}}]{Crida2006}
{Crida}, A., {Morbidelli}, A., \& {Masset}, F.~S. 2006, Icarus, 181, 587

\bibitem[{de~Val-Borro {et~al.}(2006)de~Val-Borro, Edgar, Artymowicz,
  Ciecielag, Cresswell, D'Angelo, Delgado-Donate, Dirksen, Fromang,
  Gawryszczak, Klahr, Kley, Lyra, Masset, Mellema, Nelson, Paardekooper,
  Pepliński, Pierens, Plewa, Rice, Schafer, \& Speith}]{deVal-Borro2006}
de~Val-Borro, M., Edgar, R.~G., Artymowicz, P., {et~al.} 2006, Monthly Notices
  of the Royal Astronomical Society, 370, 529

\bibitem[{{Dipierro} {et~al.}(2015){Dipierro}, {Price}, {Laibe}, {Hirsh},
  {Cerioli}, \& {Lodato}}]{DiPierro2015}
{Dipierro}, G., {Price}, D., {Laibe}, G., {et~al.} 2015, \mnras, 453, L73

\bibitem[{{Dong} {et~al.}(2015){Dong}, {Zhu}, \& {Whitney}}]{Dong2015}
{Dong}, R., {Zhu}, Z., \& {Whitney}, B. 2015, \apj, 809, 93

\bibitem[{Epstein(1923)}]{Epstein1923}
Epstein, P.~S. 1923, Physical Review, 23, 710

\bibitem[{{Espaillat} {et~al.}(2014){Espaillat}, {Muzerolle}, {Najita},
  {Andrews}, {Zhu}, {Calvet}, {Kraus}, {Hashimoto}, {Kraus}, \&
  {D'Alessio}}]{Espaillat2014}
{Espaillat}, C., {Muzerolle}, J., {Najita}, J., {et~al.} 2014, Protostars and
  Planets VI, 497

\bibitem[{{Flock} {et~al.}(2015){Flock}, {Ruge}, {Dzyurkevich}, {Henning},
  {Klahr}, \& {Wolf}}]{Flock2015}
{Flock}, M., {Ruge}, J.~P., {Dzyurkevich}, N., {et~al.} 2015, Astronomy \&
  Astrophysics, 574, A68

\bibitem[{Fouchet {et~al.}(2007)Fouchet, Maddison, \& Murray}]{Fouchet2007}
Fouchet, L., Maddison, S.~T., \& Murray, J.~R. 2007, Astronomy \& Astrophysics,
  474, 1037

\bibitem[{{Haghighipour} \& {Boss}(2003)}]{Haghighipour2003}
{Haghighipour}, N. \& {Boss}, A.~P. 2003, The Astrophysical Journal, 583, 996

\bibitem[{Kanagawa {et~al.}(2015)Kanagawa, Muto, Tanaka, Tanigawa, Takeuchi,
  Tsukagoshi, \& Momose}]{Kanagawa2015}
Kanagawa, K.~D., Muto, T., Tanaka, H., {et~al.} 2015, The Astrophysical Journal
  Letters, 806, L15

\bibitem[{{Korycansky} \& {Papaloizou}(1996)}]{Korycansky1996}
{Korycansky}, D.~G. \& {Papaloizou}, J.~C.~B. 1996, The Astrophysical Journal
  Supplement Series, 105, 181

\bibitem[{{Kwon} {et~al.}(2011){Kwon}, {Looney}, \& {Mundy}}]{Kwon2011}
{Kwon}, W., {Looney}, L.~W., \& {Mundy}, L.~G. 2011, The Astrophysical Journal,
  741, 3

\bibitem[{{Landau} \& {Lifshitz}(1959)}]{Landau1959}
{Landau}, L.~D. \& {Lifshitz}, E.~M. 1959, {Fluid mechanics}
  (Butterworth-Heinemann)

\bibitem[{Li {et~al.}(2001)Li, Colgate, Wendroff, \& Liska}]{Li2001}
Li, H., Colgate, S.~A., Wendroff, B., \& Liska, R. 2001, The Astrophysical
  Journal, 551, 874

\bibitem[{{Lin} \& {Papaloizou}(1979)}]{Lin1979}
{Lin}, D.~N.~C. \& {Papaloizou}, J.~C.~B. 1979, Monthly Notices of the Royal
  Astronomical Society, 186, 799

\bibitem[{{Lin} \& {Papaloizou}(1986)}]{Lin1986}
{Lin}, D.~N.~C. \& {Papaloizou}, J.~C.~B. 1986, The Astrophysical Journal, 309,
  846

\bibitem[{{Lin} \& {Papaloizou}(1993)}]{Lin1993}
{Lin}, D.~N.~C. \& {Papaloizou}, J.~C.~B. 1993, in Protostars and Planets III,
  ed. E.~H. {Levy} \& J.~I. {Lunine} (University of Arizona Press), 749--835

\bibitem[{Masset(2000)}]{Masset2000}
Masset, F.~S. 2000, Astronomy \& Astrophysics Supplement Series, 141, 165

\bibitem[{{Masset} {et~al.}(2006){Masset}, {D'Angelo}, \& {Kley}}]{Masset2006}
{Masset}, F.~S., {D'Angelo}, G., \& {Kley}, W. 2006, The Astrophysical Journal,
  652, 730

\bibitem[{{M\"{u}ller} {et~al.}(2012){M\"{u}ller}, {Kley}, \&
  {Meru}}]{Muller2012}
{M\"{u}ller}, T.~W.~A., {Kley}, W., \& {Meru}, F. 2012, Astronomy \&
  Astrophysics, 541, A123

\bibitem[{Nakagawa {et~al.}(1986)Nakagawa, Sekiya, \& Hayashi}]{Nakagawa1986}
Nakagawa, Y., Sekiya, M., \& Hayashi, C. 1986, Icarus, 67, 375

\bibitem[{{Paardekooper}(2007)}]{Paardekooper2007}
{Paardekooper}, S.-J. 2007, Astronomy \& Astrophysics, 462, 355

\bibitem[{{Paardekooper} \& {Mellema}(2004)}]{Paardekooper2004}
{Paardekooper}, S.-J. \& {Mellema}, G. 2004, Astronomy \& Astrophysics, 425, L9

\bibitem[{{Paardekooper} \& {Mellema}(2006)}]{Paardekooper2006}
{Paardekooper}, S.-J. \& {Mellema}, G. 2006, Astronomy \& Astrophysics, 453,
  1129

\bibitem[{{Papaloizou} \& {Lin}(1984)}]{Papaloizou1984}
{Papaloizou}, J.~C.~B. \& {Lin}, D.~N.~C. 1984, The Astrophysical Journal, 285,
  818

\bibitem[{Papaloizou \& Terquem(2006)}]{Papaloizou2006}
Papaloizou, J. C.~B. \& Terquem, C. 2006, Reports on Progress in Physics, 69,
  119

\bibitem[{{Robitaille} {et~al.}(2007){Robitaille}, {Whitney}, {Indebetouw}, \&
  {Wood}}]{Robitaille2007}
{Robitaille}, T.~P., {Whitney}, B.~A., {Indebetouw}, R., \& {Wood}, K. 2007,
  The Astrophysical Journal Supplement Series, 169, 328

\bibitem[{Supulver \& Lin(2000)}]{Supulver2000}
Supulver, K.~D. \& Lin, D. N.~C. 2000, Icarus, 146, 525

\bibitem[{{Weidenschilling}(1977)}]{Weidenschilling1977}
{Weidenschilling}, S.~J. 1977, Monthly Notices of the Royal Astronomical
  Society, 180, 57

\bibitem[{{Whipple}(1972)}]{Whipple1972}
{Whipple}, F.~L. 1972, in From Plasma to Planet, ed. A.~{Elvius}, 211

\bibitem[{Youdin(2010)}]{Youdin2010}
Youdin, A.~N. 2010, in Physics and Astrophysics of Planetary Systems, ed.
  T.~Montmerle, D.~Ehrenreich, \& A.-M. Lagrange, Vol.~41 (EAS Publications
  Series), 187--207

\bibitem[{{Zhang} {et~al.}(2015){Zhang}, {Blake}, \& {Bergin}}]{Zhang2015}
{Zhang}, K., {Blake}, G.~A., \& {Bergin}, E.~A. 2015, \apjl, 806, L7

\bibitem[{{Zhu} {et~al.}(2012){Zhu}, {Nelson}, {Dong}, {Espaillat}, \&
  {Hartmann}}]{Zhu2012}
{Zhu}, Z., {Nelson}, R.~P., {Dong}, R., {Espaillat}, C., \& {Hartmann}, L.
  2012, The Astrophysical Journal, 755, 6

\bibitem[{{Zhu} {et~al.}(2014){Zhu}, {Stone}, {Rafikov}, \& {Bai}}]{Zhu2014}
{Zhu}, Z., {Stone}, J.~M., {Rafikov}, R.~R., \& {Bai}, X.-N. 2014, The
  Astrophysical Journal, 785, 122

\end{thebibliography}

  %%%%%%%%%%%%%%%%%%%%%%
  \appendix
  %%%%%%%%%%%%%%%%%%%%%%

  %%%%%%%%%%%%%%%%%%%%%%
  \section{Disk stability}\label{sec::stability}
\begin{figure}
      \centering
      \includegraphics[width=.45\textwidth]{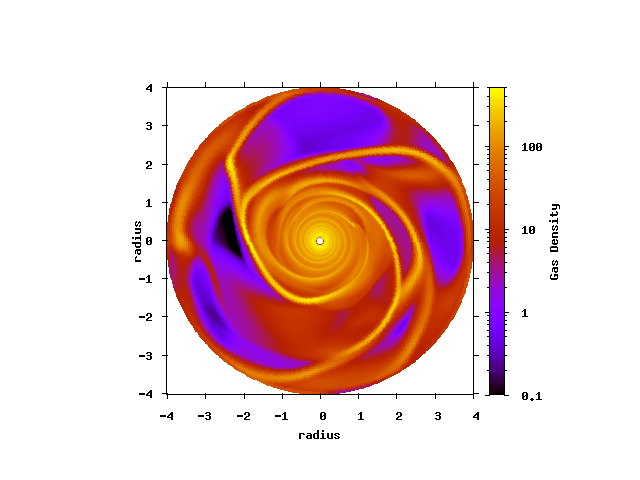}
      \includegraphics[width=.45\textwidth]{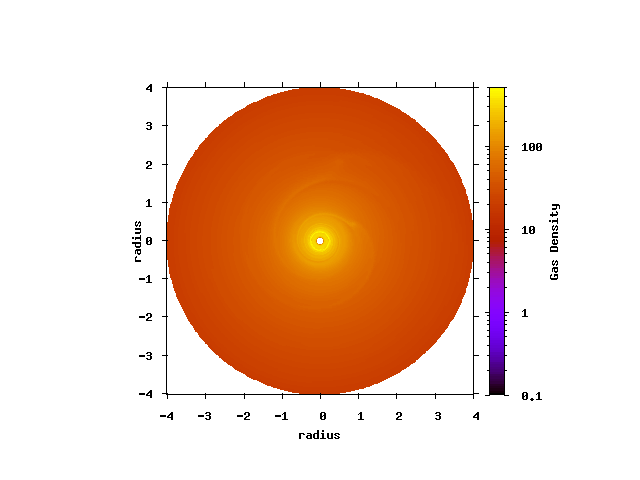}
      \caption{Disk surface density after 20 inner planetary orbits for the
      isothermal case (top panel) and fully radiative case (bottom panel) where
      the self-gravity of the disk has been considered.~\label{Fig:sg}}
    \end{figure}

  %%%%%%%%%%%%%%%%%%%%%%
  The disk parameters adopted in this work were selected in order to match the
  observational data \citep{Kwon2011}.
  We point out that the disc to star mass ratio is relatively high and,
  in the locally isothermal approximation, it is gravitationally unstable in the
  outer regions considering its Toomre parameter
  \begin{equation}
    Q = \frac{h}{r}\frac{M_\star}{M_d}\frac{2(r_{out}-r_{in})}{r} \simeq
    \frac{1.6}{r}.
  \end{equation}
  However, the isothermal equation of state is usually a poor representation of
  the temperature distribution in the disk, especially for the inner regions of
  protoplanetary disks around young stars.
  In order to validate this model we ran an additional hydro simulation in which
  we include a more realistic equation of state where the radiative transport
  and radiative cooling are included.
  From Fig.~\ref{Fig:sg} it is clear that the more appropriate equation of state
  prevent the disk to become even partially unstable.
  The different equation of state could in principle affect the gap opening
  timescale, however the long term evolution of the particle dynamics is not
  expected to vary significantly.
  Moreover, the important parameter in our study is the stopping time of the
  particles, so the results obtained remain valid for different values of the
  surface density profile, scaling accordingly the dust size, as discussed in
  Section~\ref{par:sta}.

  %%%%%%%%%%%%%%%%%%%%%%
  \section{Integrators}\label{sec::integrators}
  %%%%%%%%%%%%%%%%%%%%%%
  In order to model the dynamics of the particle population in our simulations
  we tried different integrators.

    %---------------------
    \subsection{Semi-implicit integrator in polar coordinates}
    %---------------------
    In order to follow the dynamics of particles well-coupled with the gas,
    which have a stopping time much smaller than the time step adopted to
    evolve the gas dynamics, we adopted the semi-implicit Leapfrog
    (Drift-Kick-Drift) integrator described in~\citet{Zhu2014} in polar
    coordinates.
    This method guarantees the conservation of the physical quantities for the
    long term simulations performed in this paper, and at the same time it is
    faster than an explicit method.

      \paragraph{Scheme}
      Half Drift:
      \begin{align}
        v_{\mathrm{R},n+1} &= v_{\mathrm{R},n} \\
        l_{n+1} &= l_n \\
        R_{n+1} &= R_n + v_{\mathrm{R},n}\frac{\mathrm{d}t}{2} \\
        \phi_{n+1} &= \phi_{n}+\frac{1}{2}\left(
        \frac{l_n}{R_n^2}+\frac{l_{n+1}}
        {R_{n+1}^2}\right) \frac{\mathrm{d}t}{2}
      \end{align}

      Kick:
      \begin{align}
        R_{n+2} = &\ R_{n+1} \\
        \phi_{n+2} = &\ \phi_{n+1} \\
        l_{n+2} = &\ l_{n+1} +
	      \frac{dt}{1+\frac{\mathrm{d}t}{2t_{\mathrm{s},n+1}}}
        \left[-{\left(\frac{\partial\Phi}{\partial\phi}\right)}_{n+1} +
        \frac{v_{\mathrm{g,\phi},n+1}R_{n+1}-l_{n+1}}{t_{\mathrm{s},n+1}}
        \right] \\
        v_{\mathrm{R},n+2} = &\ v_{\mathrm{R},n+1} +
        \frac{dt}{1+\frac{\mathrm{d}t}{2t_{\mathrm{s},n+1}}}
        \Bigg[
		\frac{1}{2}\Bigg(
			\frac{l_{n+1}^2}{R_{n+1}^3}+
        		\frac{l_{n+2}^2}{R_{n+2}^3}
		\Bigg)-
        	\Bigg(
			\frac{\partial\Phi}{\partial R}
		\Bigg)_{n+1} + \\
        	&\ +\frac{v_{\mathrm{g,R},n+1}-v_{\mathrm{R},n+1}}{t_{\mathrm{s},n+1}}
        \Bigg]
      \end{align}

      Half Drift:
      \begin{align}
        v_{\mathrm{R},n+3} &= v_{\mathrm{R},n+2} \\
        l_{n+3} &= l_{n+2} \\
        R_{n+3} &= R_{n+2} + v_{\mathrm{R},n+3}\frac{\mathrm{d}t}{2} \\
        \phi_{n+3} &= \phi_{n+2}+\frac{1}{2}\left(
        \frac{l_{n+2}}{R_{n+2}^2}+\frac{l_{n+3}}
        {R_{n+3}^2}\right) \frac{\mathrm{d}t}{2}
      \end{align}

      where $v_{\mathrm{R}}$ is the radial velocity, $l$ the angular
      momentum, $R$ the cylindrical radius, and $\phi$ the polar angular coordinate.
      The index $n$ shows the step at which the various quantities are
      considered.
      Further information reguarding to the integrator can be found in~\citet{Zhu2014}.

    %---------------------
    \subsection{Fully-implicit integrator in polar coordinates}
    %---------------------
    For particles with stopping time much smaller than the numerical time
    step, the drag term can dominate the gravitational force term, causing the
    numerical instability of the integrator.
    Thus, it is necessary to adopt a fully implicit integrator
    following~\citet{Bai2010a,Zhu2014}

      \paragraph{Scheme}
      Predictor step:
      \begin{align}
        R_{n+1} &= R_n + v_{\mathrm{R},n} \mathrm{d}t \\
        \phi_{n+1} &= \phi_n+\frac{l_n}{R_n^2}
        \mathrm{d}t
      \end{align}

      Shift:
      \begin{align}
        v_{\mathrm{R},n+1} &= v_{\mathrm{R},n} +
        \frac{{\diff t}/2}
        {1+\diff t{\left(
          \frac{1}{2t_{\mathrm{s},n}} +
          \frac{1}{2t_{\mathrm{s},n+1}} +
          \frac{\diff t}{2t_{\mathrm{s},n}t_{\mathrm{s},n+1}}
        \right)}}\cdot \\
        & \cdot \Bigg[
          -{\left(\frac{\partial\Phi}{\partial R}\right)}_n
          -\frac{v_{\mathrm{R},n}-v_{\mathrm{g,R},n}}{t_{\mathrm{s},n}}
          +\frac{l_n^2}{R_n^3} +
          \Bigg(
            -{\left(\frac{\partial\Phi}{\partial R}\right)}_{n+1}+
            \nonumber \\
            &-\frac{v_{\mathrm{R},n}-
            v_{\mathrm{g,R},n+1}}{t_{\mathrm{s},n+1}}
            +\frac{l_{n+1}^2}{R_{n+1}^3}
          \Bigg)
          {\left(1+\frac{\diff t}{t_{\mathrm{s},n}}\right)}
        \Bigg] \nonumber \\
        l_{n+1} &= l_{n} +
        \frac{{\diff t}/2}
        {1+\diff t\left(\frac{1}{2t_\mathrm{s,n}}+\frac{1}
        {2t_{\mathrm{s},n+1}}+
        \frac{\diff t}{2t_{\mathrm{s},n}t_{\mathrm{s},n+1}}\right)}\cdot \\
        &\cdot\Bigg[
          -{\left(\frac{\partial\Phi}{\partial \phi}\right)}_n
          -\frac{l_n - R_n v_{\mathrm{g,\phi},n}}{t_{\mathrm{s},n}}
          +\Bigg(
            -{\left(\frac{\partial\Phi}{\partial \phi}
            \right)}_{n+1} + \nonumber \\
            &-\frac{l_n - R_{n+1} v_{\mathrm{g,\phi},n+1}}
            {t_{\mathrm{s},n+1}}
          \Bigg)
          {\left(1+\frac{\diff t}{t_{\mathrm{s},n}}\right)}
        \Bigg] \nonumber
      \end{align}

      Corrector step:
      \begin{align}
        R_{n+1} &= R_n + \frac{1}{2}(v_{\mathrm{R},n}+
        v_{\mathrm{R},n+1})\mathrm{d}t \\
        \phi_{n+1} &= \phi_{n} + \frac{1}{2}\left(
        \frac{l_{n}}{R_{n}^2}+\frac{l_{n+1}}
        {R_{n+1}^2}\right) \mathrm{d}t
      \end{align}

  %%%%%%%%%%%%%%%%%%%%%%
  \section{Particle tests}\label{sec::partests}
  %%%%%%%%%%%%%%%%%%%%%%
  In order to test the numerical integrators described in Sec.~\ref{sec::integrators},
  we did an orbital test and a drift test proposed by~\citet{Zhu2014}.

    %---------------------
    \subsection{Orbital tests}
\begin{figure}
      \centering
      \includegraphics[width=.45\textwidth]{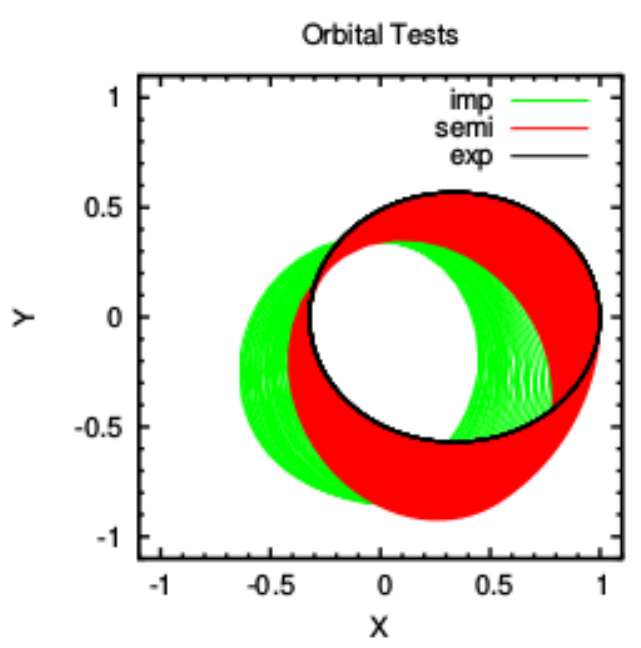}
      \caption{Orbital evolution of a dust particle released at
        $r=1$, $\phi=0$, with $v_\phi=0.7$ for the different integrators adopted
        in the simulations (red and green curves), compared to the solution from
        an explicit integrator (black curve).~\label{orbital}}
    \end{figure}

    %---------------------
    We release one dust particle at $r=1$, $\phi=0$, with $v_\phi=0.7$, and
    integrate it for $20$ orbits.
    The time-steps $\Delta t$ are varied between $0.1$ and $0.01$ in units of
    the orbital time.
    The results are shown in Figure~\ref{orbital}.
    The particle follows an eccentric orbit with $e = 0.51$.
    The time steps are $\Delta t = 0.1$, compared with the orbital time
    ($2\pi$).
    The precession observed is due to the fact that even symplectic integrators
    cannot simultaneously preserve angulat momentum and energy exactly.
    The advantage of the semi-implicit scheme is that it does preserve geometric
    properties of the orbits, while the fully implicit integrator does not.
    For comparison, the orbit is calculated also with an explicit integrator,
    but with a much smaller time step $\Delta t=0.01$, showing no visible
    precession.
    This behavior is recovered also with the implicit schemes reducing the
    timestep.
    Since $\Delta t = 0.01$ is normally the time step used in our planet-disk
    simulations, our integrators are quite accurate even if we integrate the
    orbit of particles having moderate eccentricity.

    %---------------------
    \subsection{2D drift tests}
    %---------------------
    We model a 2D gaseous disk in hydrostatic equilibrium with
    $\Sigma\propto r^{-1}$ and release particles with different stopping times
    from $r=1$.
    The radial domain is $[0.5, 3]$ with a resolution in the radial direction
    of $400$ cells.
    The drift speed at the equilibrium is given by~\citep{Nakagawa1986}
    \begin{equation}\label{eq:drifteq}
      v_\mathrm{R,d}=\frac{\tau_\mathrm{s}^{-1}v_\mathrm{R,g}-\eta
      v_\mathrm{K}}{\tau_\mathrm{s}+\tau_\mathrm{s}^{-1}}
    \end{equation}
    where $v_\mathrm{R,g}$ is the gas radial velocity, $\eta$ is the ratio
    of the gas pressure gradient to the stellar gravity in the radial
    direction, and we consider $v_\mathrm{R,g}=0$ since we are at the
    equilibrium.
    \begin{figure}
      \centering
      \includegraphics[width=.45\textwidth]{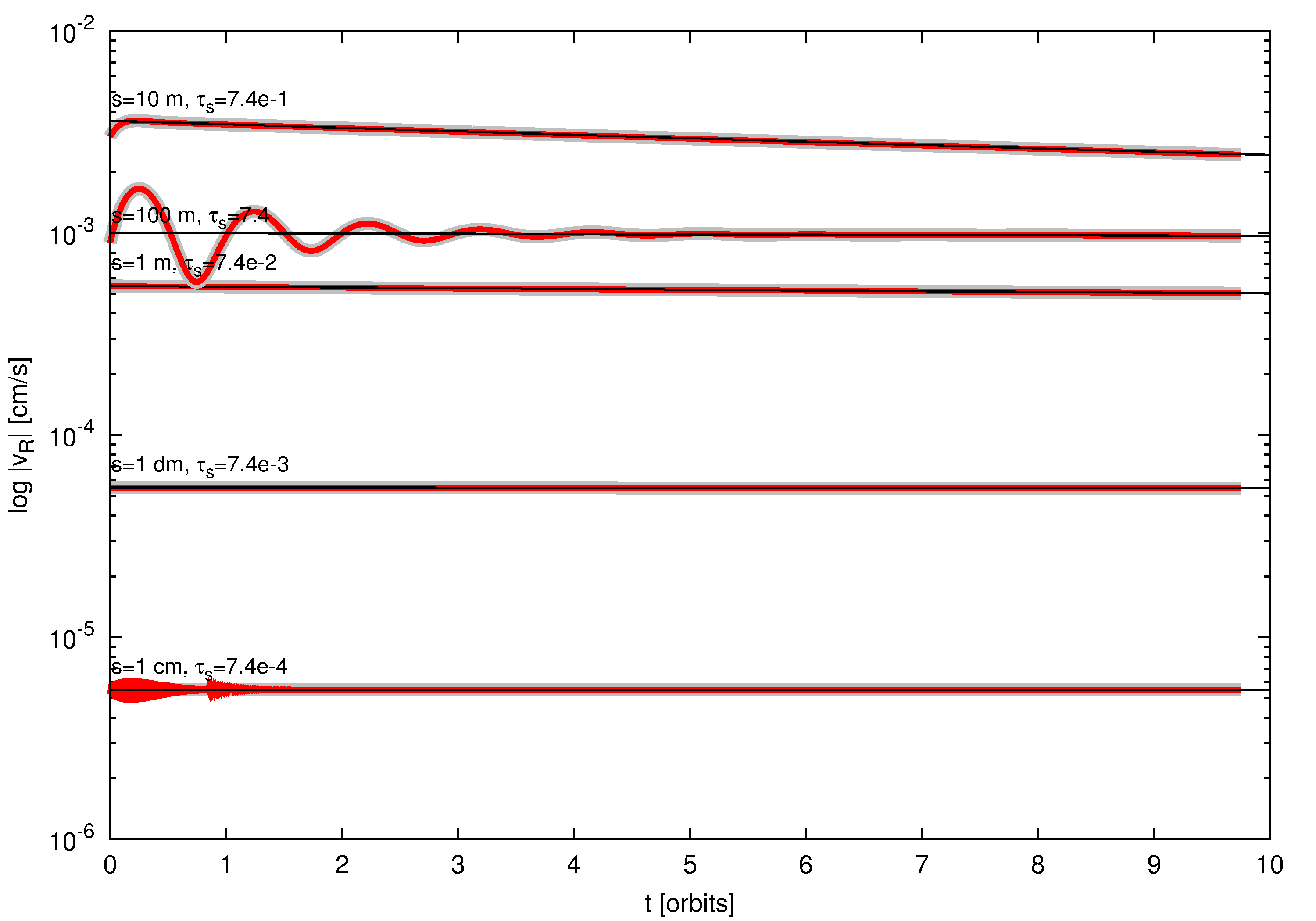}
      \caption{Evolution of the particle drift speed
        in the first 10 orbits for particles with different stopping times.
        The analytic solution obtained from
        eq.~(\ref{eq:drifteq}) is plotted with a black line, while the
        drift speed obtained from the semi-implicit and fully-implicit
        integrators are displayed with red and grey lines
        respectively.~\label{orbital2}}
    \end{figure}
    Figure~\ref{orbital2} shows the evolution of the particle radial
    velocity in the first 10 orbits for the implicit (red) and semi-implicit
    (grey) integrators together with the analytic solution (black) obtained from
    eq.~(\ref{eq:drifteq}).
    The particle drift speed reaches almost immediately the expected drift
    speed.
    The semi-implicit integrator reaches the equilibrium speed on a
    longer timescale than the fully-implicit one only for the lower stopping
    time (smaller particles).

\end{document}